# An Overview of NRL's NAUTILUS: A Combination SIMS-AMS for Spatially Resolved Trace Isotope Analysis

Evan E. Groopman*[1], David G. Willingham[1], Albert J. Fahey[2], Kenneth S. Grabowski[1]

[1]Materials Science and Technology Division, U.S. Naval Research Laboratory, Washington, DC, 20375, USA
[2]Microscopy & Surface Analysis, Corning, Inc., Corning, NY 14831, USA
*evan.groopman@nrl.navy.mil

We present a description of the capabilities and performance of the NAval Ultra-Trace Isotope Laboratory's Universal Spectrometer (NAUTILUS) at the U.S. Naval Research Laboratory. The NAUTILUS combines secondary ion mass spectrometry (SIMS) and single-stage accelerator mass spectrometry (SSAMS) into a single unified instrument for spatially resolved trace element and isotope analysis. The NAUTILUS instrument is essentially a fully functional SIMS instrument with an additional molecule-filtering detector, the SSAMS. The combination of these two techniques mitigates the drawbacks of each and enables new measurement paradigms for SIMS-like microanalysis. Highlighted capabilities include molecule-free raster ion imaging for direct spatially resolved analysis of heterogeneous materials with or without perturbed isotopic compositions. The NAUTILUS' sensitivity to trace elements is at least 10× better than commercial SIMS instruments due to near-zero background conditions. We describe the design and construction of the NAUTILUS, and its performance applied to topics in nuclear materials analysis, cosmochemistry, and geochemistry.

## Introduction

Accelerator mass spectrometry (AMS) is synonymous with ultra-trace isotope analysis, while secondary ion mass spectrometry (SIMS) is the premier spatially resolved, sensitive, surface analysis technique. We have successfully designed and built the NAval Ultra-Trace Isotope Laboratory's Universal Spectrometer (NAUTILUS) at the U.S. Naval Research Laboratory (NRL)[4-9], which combines the modified hardware from two commercial instruments, an Ametek Cameca ims 4f SIMS[10] and a National Electrostatics Corporation (NEC) single-stage AMS (SSAMS)[11-13], together with custom control hardware and software. The motivation for this novel combination MS-MS instrument is to utilize the aforementioned advantages of each technique in a manner which simultaneously mitigates each technique's drawbacks. While SIMS maintains excellent sensitivity for materials analysis with down to micrometer spatial and nanometer depth resolution, the sputtering process produces molecular secondary ions. These may interfere at the same mass-to-charge ratio (m/z) with isotopes of interest, especially for high-mass and/or trace analyses. The trade-off between increased mass resolving power (MRP), typically defined as the full width at 10% peak height, and decreased instrumental transmission can make high-mass analyses impractical, especially for trace isotopes (e.g., detection of $^{236}U$ in the presence of $^{235}U^1H)^{[14]}$. AMS excels at removing molecular isobaric interferences, but these instruments typically analyze bulk samples without spatial resolution either due to chemical sample preparation or large sputter source size. While SIMS may analyze positive or negative secondary ions, tandem AMS instruments are limited to injecting negative ions. This dramatically decreases their sensitivity to electropositive elements, since molecular ions such as $FeO^-$ or $UO^-$ must be generated to transport the element of interest, Fe or U, into the AMS. This orders-of-magnitude decrease in sensitivity may be prohibitive for small-sample analyses, such as those performed by SIMS. The SSAMS, with a smaller footprint and lower energy than most tandem AMS instruments, can accept positive or negative ions, making it the ideal AMS system to integrate with a SIMS. We focus on electropositive element analyses with the NAUTILUS, since most high-mass elements and much of the periodic table are in this category.

The combination of SIMS and SSAMS enables new measurement paradigms for SIMS-type analyses of materials. In particular, novel capabilities such as direct raster ion imaging allows for elemental and isotopic heterogeneities of trace elements to be identified within complex matrices. The sensitivity of measurements can be additionally increased through the use of novel ion beams and sample flooding gases. Ordinarily these do not increase the overall signal-to-noise (SNR) ratio of a measurement since they often enhance the magnitude and complexity of the molecular background in addition to enhancing the signal of interest. The capabilities of the NAUTILUS have been demonstrated through the discovery of fissionogenic Cs and Ba captured in Ru metal/sulfide aggregates within Oklo reactor materials[5]; direct measurement and identification of $^{236}$U in $U_3O_8$ particles covered with monazite "dirt"[4]; and through direct measurement of trace rare earth elements (REEs) in hibonite and other minerals where the molecular background was $(10^4$-$10^5)\times$ higher than the signals of interest[6]. What follows is a description of the NAUTILUS instrument and its performance. We describe the novel capabilities developed and enabled by the combination SIMS-SSAMS, the extensive modifications made to the Cameca ims 4f and NEC SSAMS, the control hardware and software, and future developments.

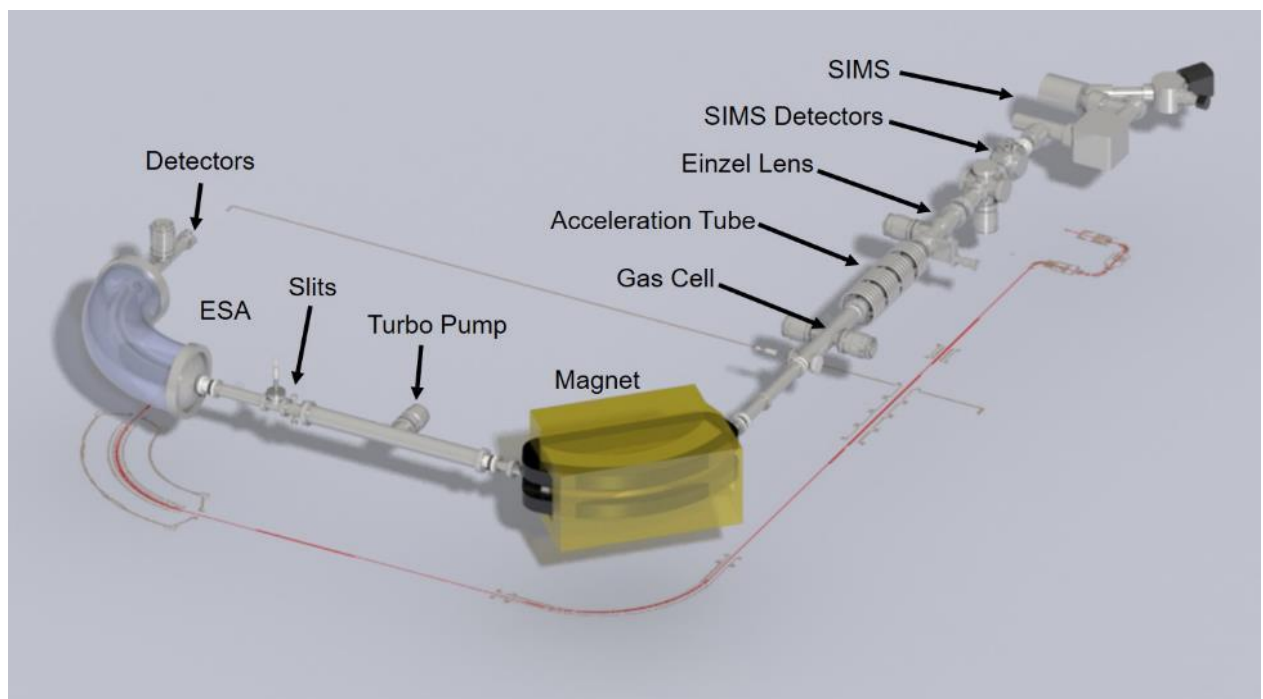

Figure 1: 3D CAD model of the NAUTILUS with SIMION cutaway view. The NAUTILUS is a standalone SIMS instrument with a single-stage accelerator mass spectrometer "detector" that eliminates molecular isobaric interferences.

## NAUTILUS Overview

The NAUTILUS at NRL combines two modified commercial instruments, a Cameca ims 4f and an NEC SSAMS, into one unified instrument. The NAUTILUS combines each technique's advantages and mitigates their drawbacks with the express purpose of performing spatially resolved trace isotope analyses free from molecular isobaric interferences. A key feature of this setup is that neither individual instrument is compromised by the unification – the SIMS remains operational as a standalone instrument and the SSAMS retains the ability for alternative ion sources to be used in lieu of the SIMS frontend. That said, the NAUTILUS can best be thought of as a SIMS instrument with a large and specialized molecule-filtering detector. Figure 1 shows a 3D computer aided design (CAD) rendering of the NAUTILUS overtop of a Monte Carlo ion optics trace generated using SIMION[15]. The CAD and SIMION models are based upon schematics provided by NEC and Cameca; the SIMION model of the ims 4f is based upon the model in Lorincik et al.[16], which does not include the primary ion optics, with the addition of a physical magnet modelled for this work. Figure 2 shows the top-down (panel A) and potential energy (panel B) SIMION views of the NAUTILUS. A Faraday cage, shown in black in panel B of Figure 2, surrounds the floated accelerator deck, but was omitted for visual clarity from Figure 1. The SSAMS itself is relatively simple in terms of its ion optical components, consisting of an acceleration tube, magnetic sector with floating flight tube, and electrostatic spherical analyzer (ESA), making it straightforward to use as a "detector" for the SIMS.

With the relative simplicity of the SSAMS design and ease of tuning, the majority of day-to-day operations involve getting ions into the accelerator, i.e. on the SIMS side. This is a critical point, as the operational characteristics of the NAUTILUS should not be conflated with more traditional AMS systems. Because the NAUTILUS is designed for spatially resolved trace analysis of materials on the µm to hundreds-of-µm scale, the quantities of material consumed and samples/applications of interest are vastly different than traditional AMS. The dynamic range of NAUTILUS measurements is similar to SIMS and quite dissimilar to traditional AMS. Therefore, the NAUTILUS can be thought of as a SIMS-based, rather than an AMS-based, technique. However, the measurement paradigm on NAUTILUS differs from SIMS on many key points due to the use of the additional mass spectrometer and molecule filter. Notably, increased MRP and/or energy filtering are no longer required to increase the SNR of an analysis suffering from molecular isobaric interferences, as in e.g., [17, 18]. These typical procedures in fact only harm abundance sensitivity by reducing ion transmission through the SIMS. We have found that the most stable and sensitive operating paradigm involves the SIMS being nearly wide-open (low MRP) to maximize transmission.

We acquired a complete Cameca ims 4f from the U.S. Naval Surface Warfare Center in Crane, IN. Subsequently, We connected the Cameca ims 4f to a NEC ±300 kV SSAMS. Modifications specific to each commercial component are detailed in later sections. The original electronics for both systems were replaced for unified control from a single computer. Use of a legacy SIMS instrument was beneficial towards the development of a prototype combination SIMS-SSAMS. As the unified system was brought online, it was significantly easier to modify, bypass, and/or control the original equipment manufacturer (OEM) analog electronics so we could implement incremental modifications. It would have been difficult/prohibitive to do the same for the integrated digital systems of current Cameca SIMS instruments. A next-generation NAUTILUS could be designed from the beginning with fully digital control in mind.

The NRL SSAMS operates at ±300 kV and was designed by NEC as a larger-geometry (~8m on a side) version of their radiocarbon SSAMS, able to analyze ions up to 300 m/z. Positive or negative ions may be injected into the SSAMS, though we concentrate primarily on analyzing electropositive elements, since these make up the majority of the periodic table, including the actinides and rare earth elements (Figure 3). Generally, elements shaded green and yellow in Figure 3 preferentially produce positive ions, those in red produce mostly negative ions, and those in orange can do both. Our sensitivity to electropositive elements is increased by several orders of magnitude relative to instruments that can only analyze negative molecular ions of these elements, usually monoxides, which allows small micrometer-sized volumes of material to be analyzed by SIMS-SSAMS. Without the ability to inject positive ions into the accelerator, much larger volumes of material would be required, as is the case for actinide measurements on the ETH TANDY[19] and CNA AMS facility[20], or sensitivity would be limited to the mostly electronegative elements, as on the UCLA MegaSIMS[21] measuring O[22].

Ion optical coupling between the SIMS and the SSAMS is achieved through use of the SIMS projection lenses and an additional Einzel lens immediately before the acceleration electrodes (Figure 2). The ion beam from the SIMS is focused so that the beam forms a waist at the center of the gas stripping tube, which is the object location for the SSAMS magnet and ESA. The high-voltage "deck" of the SSAMS is biased to -300 kV for positive ion analyses by a resistor/capacitor stack provided by NEC; a separate inverse stack was provided for analysis of negative ions. A resistor chain across 44 electrodes to Earth ground provides the acceleration bias and focusing for the SSAMS. Currently the SIMS operates at a 4.5 kV extraction. In order to extract ions in the 8 – 10 kV range, which would result in higher SIMS transmission, several resistors in the chain would need to be changed to accommodate the resulting changes in the focal properties of the acceleration tube.

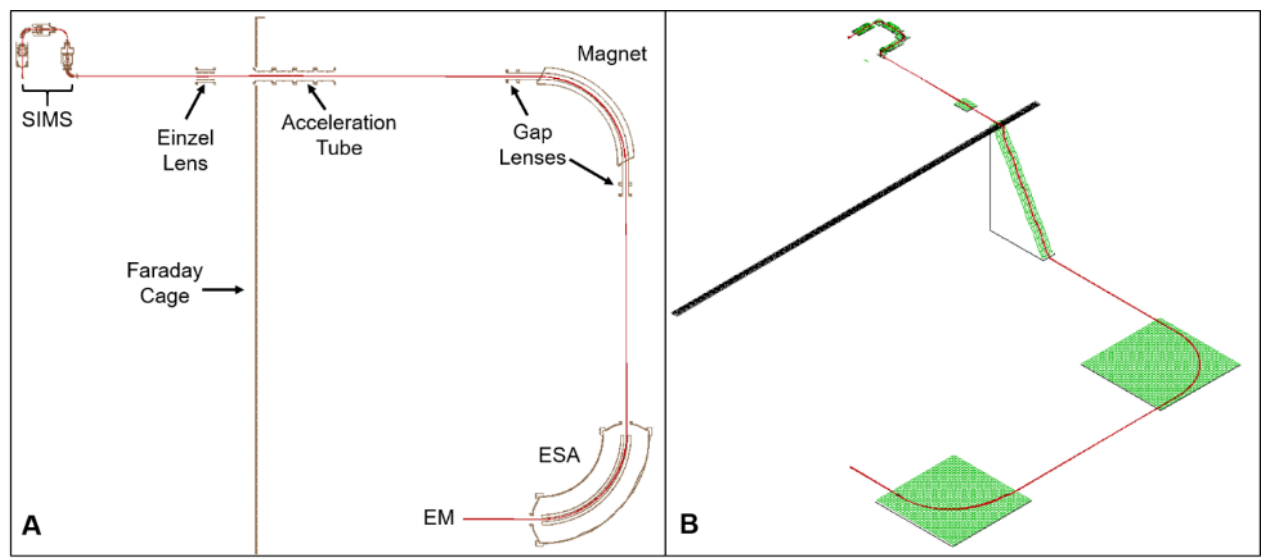

Figure 2: SIMION models of NAUTILUS. Panel A shows top-down, cutaway view of ion optical elements. The SSAMS is relatively simple in ion optical terms, consisting of an acceleration tube, magnetic sector cut by a floating flight tube, and an ESA. Panel B shows the potential energy view with the accelerator biased to -300 kV. The black stripe in panel B is the grounded Faraday cage around the accelerator.

During tuning and analysis, an ion beam with a single m/z ratio is injected serially from the SIMS into the accelerator. Helmholtz coils and deflectors in the coupling section between the SIMS and SSAMS help correct for stray magnetic fields so that ions of all masses are injected into the SSAMS at the same location and with the same angular dispersion. A gas stripping cell consisting of concentric, differentially pumped cylinders typically is filled with gas controlled by a mass flow controller. To date we have predominantly used Ar as the stripping gas. The Ar is stored at ground level in a 300 cu. ft. cylinder and is supplied to the deck by a ¼ inch polyethylene tube at 60 psi, fed through the corona rings of the acceleration tube. This pressure can insulate the 300 kV bias over the ~1 m acceleration distance. A manifold on the deck allows us to alternatively connect lecture bottles with different gas species. A regulator on the SSAMS reduces the pressure to 10 psi supplying an Alicat mass-flow controller, which controls the gas thickness in the stripping tube. For Ar, typical gas flow rates are 0.075 – 0.3 standard cubic centimeter per minute (sccm), or up to $1.6 \times 10^{-8}$ mol/cm$^2$ [6, 7]. We use a 0-1 sccm mass-flow controller for heavier gases, such as Ar, and a 0-10 sccm controller for He, which is fed at 100 psi from ground. There remains much to explore regarding the optimal gas composition and pressure for a specific analysis, since relatively little is known about gas stripping properties at "low" AMS energies, especially for high-mass elements, e.g., most elements other than C. The maximum acceptance angle of the mass spectrometer is 13.5 mrad half-angle, which accounts for some transmission loss due to scattering. The majority of transmission loss appears to be due to charge state change of the ions, especially conversion into neutrals, which cannot be easily measured. We focus solely on analyzing charge state +1 ions from the SIMS and SSAMS stripping cell, because this is where the fewest potential "look-alike" isobars (molecular or atomic species with different charge states that appear at the same mass-to-charge ratio) may be present and is the most intense measureable signal over the range of gas pressure that we employ.

The SIMS and SSAMS each contain a suite of detectors including one or more electron multipliers (EM), Faraday cups (FC), and micro channel-plate (MCP) beam imagers. Electrostatic deflectors are used to switch between detectors on the SIMS and SSAMS during tuning and data acquisition, allowing for hybrid SIMS and SSAMS measurements in a single analysis. Both the individual SIMS and combined SIMS-SSAMS instruments measure isotopes serially on a given detector. Discussion of EM electronics and use on a 300 kV accelerator is presented in later sections. Use of gas stripping additionally allows for two modes of analysis on the accelerator: molecule filtering mode, and fragment analysis mode. A more detailed discussion of these two modes follows in a later section. We run the ims 4f tuned to a low MRP between 300 and 500 for high transmission – for the latter, the ratio of exits slit to entrance slit widths is 2:1. The NAUTILUS only requires nominal mass resolution before and after molecular dissociation. Nuclear isobars above mass 40 are effectively

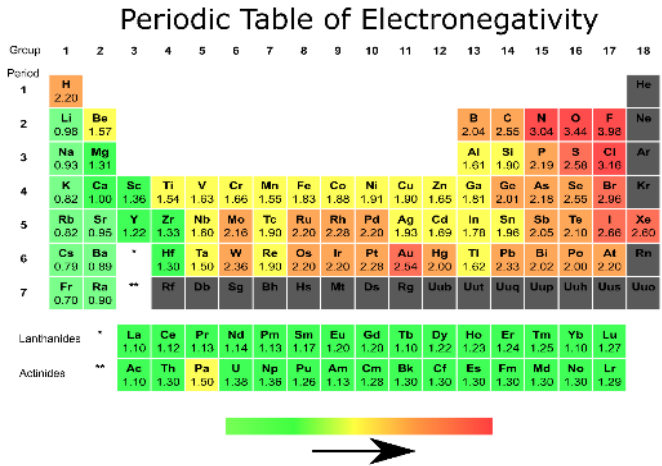

## Periodic Table of Electronegativity

Figure 3: Periodic table of electronegativity. Generally, elements shaded green and yellow preferentially produce positive ions, those in red produce mostly negative ions, and those in orange can go both ways. The SSAMS can accept positive or negative ions from the SIMS, however we focus on electropositive element analysis since these make up the majority of the periodic table and include the actinides and rare earth elements. Sensitivity is increased by several orders of magnitude when analyzing these elements as positive ions, allowing small micrometer-sized volumes of material to be analyzed by SIMS-SSAMS. Data from Rumble, Lide, & Bruno (2018) [3] and references therein.

unresolvable for trace elements. For example, at best $^{48}$Ti and $^{48}$Ca require MRP > 11,000, while $^{54}$Cr and $^{54}$Fe require MRP > 73,000, and $^{142}$Ce and $^{142}$Nd require MRP > 78,000[23]. Therefore, since the stripping gas removes molecular isobaric interferences universally, our sensitivity is maximized by operating the SIMS nearly wide-open. A significant benefit of operating the SIMS with low MRP is that the instrument is highly stable and settings are reproducible day-to-day as fluctuations in temperature, control voltages, etc., have relatively little effect on the mass spectrometer. We also open the SIMS energy slit to a bandpass of 150 eV to increase transmission relative to a typical 50 eV setting. The energy spread introduced by ion-gas collisions is on the order of 1-2 keV under typical operating conditions, so the increased incident energy dispersion has little detrimental effect downstream.

Two motor-generator pairs with insulating drive shafts provide power to the SSAMS when energized; one pair powers the vacuum system and the other powers most high-voltage equipment. A pair of monopolar Heinzinger electronic GmbH (Rosenheim, Germany) PCU 50V/100A power supplies provide power to the 7.5-ton soft iron SSAMS magnet, which requires >8 kW of power (~90 A) to set the magnetic field for U (m/z = 238). Cooling to the coils is provided by water pumped at 3 gallons per minute from a reservoir on the SSAMS, which is connected by a set of heat exchangers to the building chilled water supply at ground. Low odor base solvent, which can hold off the 300 kV deck bias, is pumped between the heat exchangers on the SSAMS and at ground.

The flight tube through the accelerator's magnetic sector is electrically isolated via two gap lenses. The flight tube is biased using a 20 kV bipolar Trek, Inc. (Lockport, NY, USA) power supply, which enables switching between masses of interest with a static magnetic field, referred to as "bouncing" in the accelerator community[24] and electrostatic peak switching (EPS) in the SIMS community[25, 26]. For a given radial flight path through the magnetic sector, the mass × energy product is conserved, so mass switching is possible by biasing the flight tube with a voltage, which changes the energy of the ion inside of the magnetic sector and therefore its radius of curvature. On the SSAMS, the range of the switching is limited to the flight tube bias relative to the ion energy: ±20 kV / 304.5 kV ≈ ±6.5%. The gap lenses and flight tube act as an Einzel lens of non-ideal geometry[6], which affects the focal properties of the beam. Within the ±6.5% switching range these effects are minor and we do not observe any fractionation. Significant benefits of EPS over magnetic switching include: rapid switching (< 0.1 s), high duty cycle, and no magnet hysteresis effects. Without hysteresis effects over the local ±6.5% switching range of the EPS, mass order does not matter during analysis. For instance, when performing U-Th-Pb measurements for radiometric dating, SIMS analyses often require an intense molecular peak for centering low-abundance $^{204}Pb^+$ at a lower mass (e.g., $^{196}[^{90}Zr_2{}^{16}O]^+$ in zircon[27]) since isotopes are magnetically cycled in mass order. Since there are no hysteresis effects due to mass analysis order on the NAUTILUS, we are not limited to cycling in mass order; therefore, $^{206}Pb^+$ may be analyzed first to center $^{204}Pb^+$. For masses separated by more than ±6.5% of the current mass, the magnetic field is switched, which results in the same types of hysteresis effects present on other SIMS instruments. In order to synchronize the magnets of both SIMS and SSAMS instruments for a given field, we modified our ims 4f to use EPS[6]. Synchronization between the two magnets enables molecule-free mass scans across the local ±6.5% EPS range (see alter sections).

Due to energy loss from collisions in the gas cell, mass peaks on the accelerator are not flat topped, however this does not affect our analytical precision. Figure 4 shows a comparison of an EPS scan on the SSAMS for $^{184}W^+$ with and without gas. With stripping gas there is a shoulder on the high-mass (energy) side of the peak, while the low-mass side of the peak is raised to higher intensity, followed by a low-energy tail. By contrast, the peak scan with no gas is nicely flat-topped. There is a small low-energy scattering tail with gas flow set to zero since the mass flow controller diaphragm does not completely close and this particular unit does not have a positive shutoff valve. We plan to address these limitations in future revisions of the NAUTILUS. Despite not being flat-topped with gas, the peaks are fairly broad and slope gently about their center-of-mass (COM). Since we operate with low MRP, the beam size is considerably smaller than the slit width between the SSAMS magnet and the ESA. During acquisition, peak centering is performed using EPS serially on both the SIMS and SSAMS by measuring a 7-point COM spanning there is 1.5× the full-width at half-max (FWHM) of the peak. The specific peak shape depends upon the flow rate of the gas and therefore the energy-loss distribution, however, for a given flow rate during acquisition or across the length of a day the shape is completely stable.

**Molecule Filtering Mode**

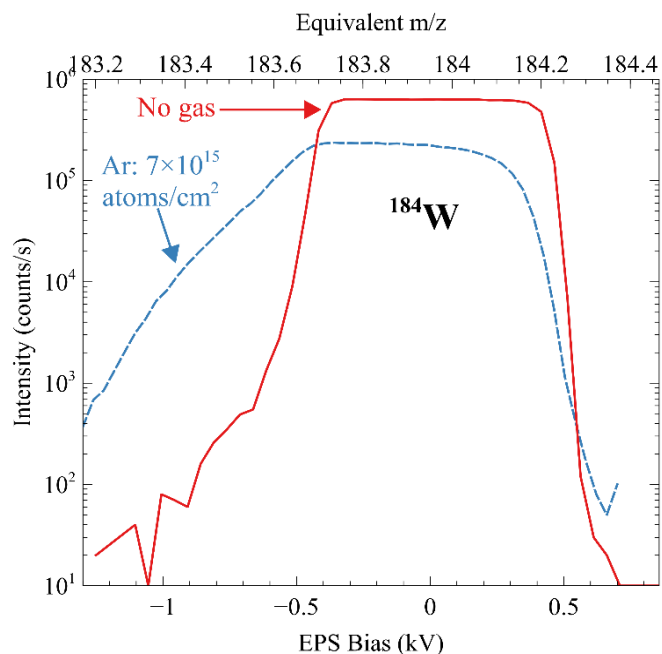

Figure 4: Mass spectrum of $^{184}$W collected with and without stripping gas by scanning the SSAMS EPS across the slits (3 mm wide) between the magnet and ESA. A small leak of gas through the mass flow controller diaphragm causes a minor scattering tail at low energy for each peak. The peak shape without gas is flat-topped, even at low MRP. With gas, the peak is not perfectly flat topped, but the shape is stable for a specific gas flow rate, so this does not affect our precision or accuracy.

The most common analysis mode on the NAUTILUS utilizes the SSAMS as a molecule filter, where the SIMS and SSAMS magnets are set to the same nominal mass and ions of a single m/z ratio are injected from the former into the latter. Following molecular dissociation in the gas cell, the SSAMS magnet filters the molecular fragments from the clean atomic signal, which is deflected into one of the detectors. Coulomb interactions between the electron clouds of beam ions and stripping gas atoms result in energy loss and charge transfer from the beam. Coulomb interactions between their screened nuclei cause scattering and energy dispersion. These combined effects vary by element. The energy dispersion is corrected in part by the SSAMS ESA when using either the end-station EM or MCP detectors. To first order, the NAUTILUS' SNR, e.g., the transmission ratio of atomic and molecular ion species, is modulated by the gas flow rate through the stripping tube. This is analogous to MRP. In both cases, increased sensitivity to a specific isotope often comes at the expense of transmission. For both, there exists an optimal condition where specificity is high and wastage is minimized, although the exact parameters depend upon the signals of interest and the sample matrix. For high-mass elements, such as the actinides, or elements where the molecular background is often larger than the signals of interest, such as the rare earths, molecular dissociation leads to a higher SNR than high MRP would alone. This is especially important for samples where molecular isobars are unresolvable, e.g., $^{235}U^1H^+$ and $^{236}U^+$ (MRP > 38,000). It is also crucial in samples where isotopic abundances are perturbed from normal, so corrections cannot be made based upon a measured $^1MX^+/^iM^+$ ratio without knowing the isotopic abundance of $^1M$ *a priori*[17, 18].

### Molecule Fragment Mode

We also routinely use a second operation paradigm called "fragment mode", where molecular ions are purposefully injected into the SSAMS, dissociated, and specific fragments are analyzed. This is similar to how actinide measurements on tandem AMS instruments are performed, e.g., Christl et al.[19, 20]. There are several benefits to this mode of operation:

1) Several elements have larger molecular oxide ion yields vis-à-vis their atomic yields (e.g., $UO^+/U^+ > 1$) and will have increased overall sensitivity.

2) Fragment mode requires less gas flow and therefore less scattering loss, increasing SNR.

3) H or other light elements may be measured by proxy as part of molecules.

4) Potential interfering species may be identified by their fragment spectra.

The first benefit is straightforward, since higher ion yield should lead to better sensitivity and precision. It has been hypothesized, for instance, that measuring $UO^+$ ions from $U_3O_8$ particles increases measurement precision, however typical SIMS cannot resolve the interference of $^{238}U^{12}C^+$ on $^{234}U^{16}O^+$, which is common for particles on vitreous carbon planchettes. As described below, the fragments $^{238}U^+$ and $^{234}U^+$ from $^{238}U^{12}C^+$ and $^{234}U^{16}O^+$, respectively, are easily resolved on the SSAMS, so this is not a limitation on the NAUTILUS. Unlike filtering mode, where molecule reduction of anywhere between 3 to 7 orders of magnitude may be required for a given analysis, fragment mode maximizes transmission by using a smaller gas thickness. This may only result in a 2 order of magnitude reduction in molecules, but it maximizes the SNR by reducing scattering and charge state change losses. Instead of trying to remove the background at a given mass, fragment mode converts a molecular ion into the atomic ion of interest. Molecular dissociation and scattering loss vary exponentially with gas thickness, so it is optimal to only dissociate (e.g., 99% of molecules into fragments instead of 99.999%) since the last 1% increase in signal will cost significantly more in scattering loss at higher flow rates. In fragment mode, the SSAMS is tuned to a lower m/z than the incident ions, so molecules that were not dissociated would not interfere with the resulting signal, and were lost from the analysis. These two benefits combine to offer up to 20× better sensitivity than filtering mode.

Hydrogen is often difficult to measure by SIMS, not least of which because it is strongly affected by stray magnetic fields and is easily fractionated in non-uniform electric fields. Due to the long path length of the NAUTILUS, a conventional H measurement would unfeasible. However, using fragment mode, hydride molecules were injected from the SIMS, where the higher molecule mass made the ion less susceptible to stray magnetic an non-uniform electric fields, into the SSAMS where the more-massive fragment was measured as a proxy for H. Molecular secondary ions have a narrower energy distribution from the sputtering process than atomic ions, so there may be less instrumental mass fractionation. Conversely, sensitivity to H may suffer by relying on creation of a hydride molecule. We have not devoted significant time to investigating this measurement paradigm, though all other considerations for measuring volatiles by SIMS still apply to the 4f frontend of the NAUTILUS, such as the quality of the high vacuum.

Molecule fragments are energy partitioned according to the mass fraction of the original molecule. For example, an incident $^{238}U^{16}O^+$ ion with 304.5 keV of energy would dissociate into two fragments, $^{238}U^{m+}$ and $^{16}O^{n+}$, where m,n = 0,1,2, 3… for the full charge state distributions. Ions of $^{238}U$ would have a kinetic energy of 238/(238 + 16) ≈ 93.7% of the 304.5 keV (~285.3 keV), and $^{16}O$ ions would have 16/(238 + 16) ≈ 6.3%. The mass × energy product through a magnetic sector is conserved, which is the underlying principle of EPS[6, 25, 26]. It is therefore simple to calculate the effective masses of the fragment ions from the energy partitioning were they still at 304.5 keV. In the previous example, $^{238}U^{m+}$ ions would have roughly 93.7% of the molecule's kinetic energy, or an effective mass-to-charge ratio of 93.7% × 238/z ≈ 223/z. We call this 93.7% figure the mass-energy gain. In both analysis modes, ions lose an additional 1 − 2 keV due to scattering in the gas cell. Tuning the SSAMS to a specific fragment, therefore, involves a simple calculation based upon the mass of the incident molecule, the charge state of interest, and the magnetic sector's field strength and EPS voltage, which are interchangeable within ±6.5% of the sector's central mass. Since ion fragments have less energy than the incident ions, ion optical components downstream of the gas cell must all have their voltages multiplied by the mass-energy gain, including the EPS flight tube, the SSAMS ESA, and the end-station deflector which selects the EM or MCP. The NAUTILUS control software calculates and applies this gain factor automatically.

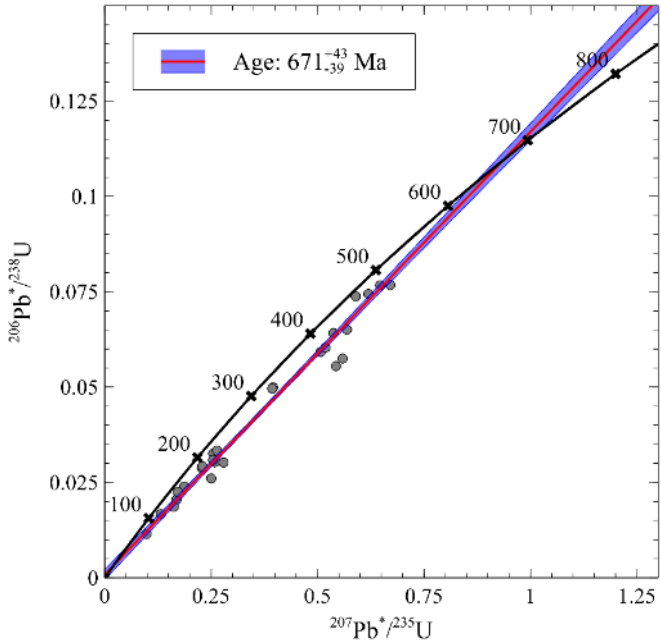

Figure 5: Example of U-Pb dating from uraninite minerals from reactor zone 13 of the Oklo natural nuclear reactor showing a discordia age of 671 ± 43/39 Ma, in agreement with previous work[2]. Uranium-lead dating is an example of using combined filter and fragment mode analyses on the NAUTILUS. Grenville skarn titanite provided by Allen Kennedy used as an age reference[1]. Oklo sample provided by Francois Gauthier-Lafaye, Maurice Pagel, and Alex Meshik.

### Combined Filtering and Fragment Modes: The Case of U-Th-Pb Dating

Combining filtering and fragment modes in a single analysis was straightforward, since the NAUTILUS measures isotopes serially. This flexibility is similar to interleaving measurements on different detectors, such as between the SIMS EM and the SSAMS EM. The only parameters required prior to analysis are the EPS center voltages for each peak (and peak widths for centering), plus a calculated mass-energy gain factor for each molecule fragment (1 for atomic ions).

The 7.5 ton soft-iron SSAMS magnet does not switch rapidly, but is quite stable once set at a given field. Settling times may be anywhere from 5 to 20 seconds, which would not be conducive to jumping during a single analysis. At high mass, ±6.5% is large enough to cover several elements, however not enough to span Pb to U for a single magnetic field setting. A combination of filtering and fragment modes in a single analysis resolves this issue, however. By setting the SSAMS magnet to a field around m/z = 214, the ±6.5% EPS range spans a mass range of 200 to 227 m/z, which fully encompasses $Pb^+$ ions injected from the SIMS and $U^+$ molecular fragments from $UO^+$ injected from the SIMS ($^{238}U^+$ from $^{238}UO^+$ has an effective mass of ~223 u). This is also true for $U^+$ fragments from injected $UO_2^+$. Therefore the SSAMS magnet can be set to one field for the entirety of the measurement, while the SIMS magnet is jumped between $Pb^+$ and $UO^+$ (or $UO_2^+$), with individual Pb and UO isotopes/fragments being centered by EPS. A demonstration of U-Pb chronometry from uraninite minerals from the reactor zone 13 of the Oklo natural nuclear reactor is shown in Figure 5. Lead isotopes in this sample from reactor zone 13[5, 28-30] are discordant due to nearby volcanism roughly 671 Ma ago, in good agreement with earlier work[2]. A sample of titanite from a Grenville skarn (OLT-1) was used as an age reference[1]. For this measurement, $^{206}Pb^+$ was used as a centering reference for $^{204}Pb^+$; all other Pb and U isotopes were centered upon individually. After sputter cleaning of a large area, there is very little common Pb in the sample.

This example further demonstrates a novel method for combining different analysis types. As described previously, $UO^+$ ions typically have a larger secondary yield than $U^+$, however the yield of $PbO^+/Pb^+ < 1$. By combining the two analysis modes, we maximize the SNR for each element in addition to achieving a more rapid analysis paradigm by keeping the SSAMS magnet at a fixed magnetic field.

**Small-geometry SIMS mode**

One of the chief features of the NAUTILUS is that it retains a complete, working small-geometry SIMS instrument, which may be operated without the SSAMS "detector". Outside of the instrument control hardware and software, the primary difference between a commercial ims 4f and our own is the physical location of the EM and FC detectors, which are housed off-axis in a detector cube, which takes the place of the OEM detector assembly. Each detector is selected electrostatically by a pair of deflector plates, while the on-axis flight path leads to the SSAMS. The SIMS projector lenses were used to correct for the longer path length to the detectors relative to their original locations. All other tuning and operation is directly analogous to other small geometry SIMS instruments.

**"Large-Geometry SIMS" Mode**

A fourth, though little used mode, is to use the full NAUTILUS instrument as a "large-geometry" SIMS instrument. In this mode, the 4f is operated nearly wide-open as in filtering and fragment modes, but no stripping gas is used. The SSAMS magnet and ESA, both 1 m radii, are in reverse geometry with moveable slits in between. In this mode the SIMS is used as an energy filter, while the high MRP is achieved on the SSAMS. The slits on the SSAMS are currently manually operated, but could be motorized if significant benefit were identified to using this mode. As it stands, tuning for high MRP would require raising and lowering the accelerator bias voltage to adjust the slits, which is cumbersome. Typically the slits are left at 3 mm spacing for low MRP in filtering and fragment modes.

**Gas Stripping and Molecular Destruction**

**Stripping Gases**

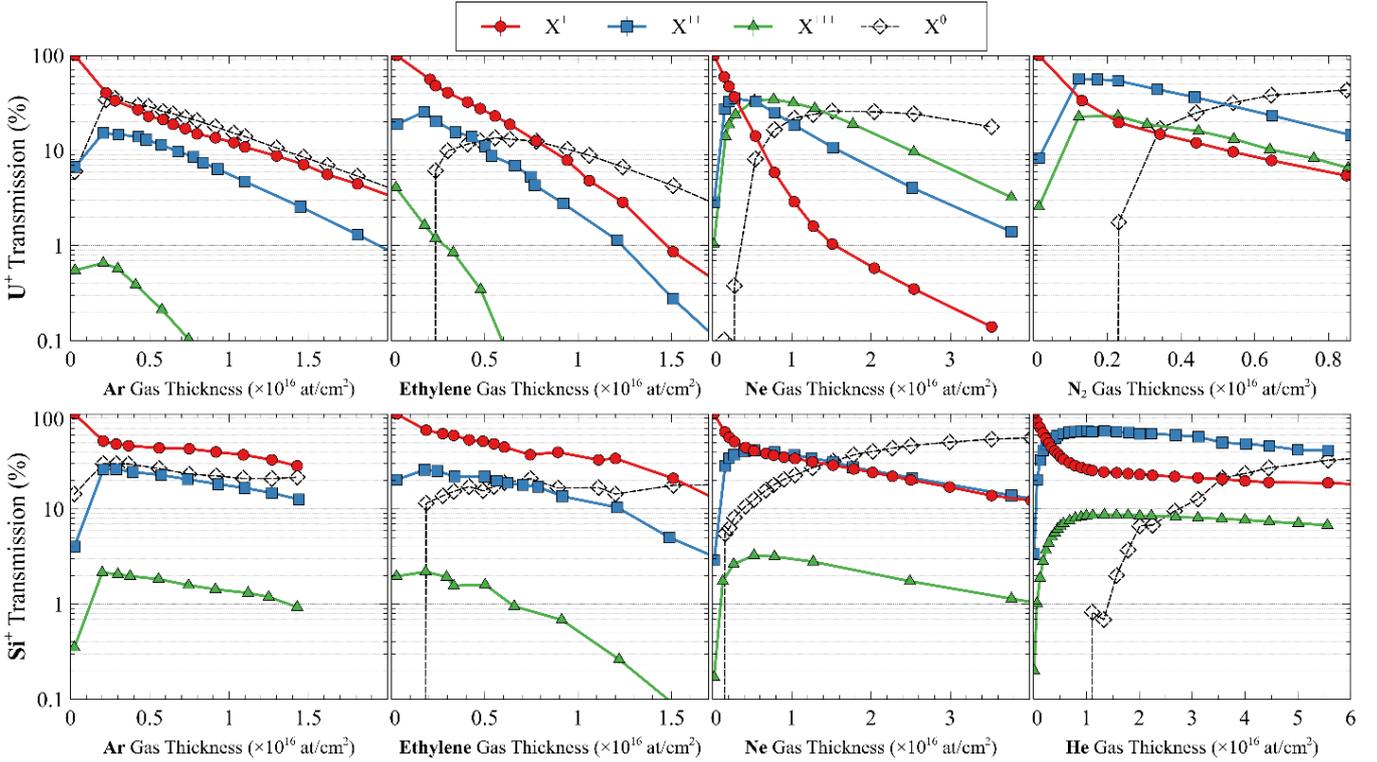

Figure 6: Comparison of charge state distributions for U$^+$ and Si$^+$ ions injected through several stripping gases. Argon and ethylene are good at producing predominantly charge +1 ions, while Ne, N$_2$, and He produce more multiply charged ions. Neutral fractions were calculated from the difference between TRIM transmission (scattering loss) and the summed charge state distributions.

To date, we predominantly used Ar gas on the NAUTILUS to perform molecule filtering and fragmentation. A mass-flow controller fed Ar into the differentially pumped gas cell, which yielded an approximately uniformly dense region of gas through which the ion beam transited. The differential pumping setup maintained this uniform gas thickness across the stripping canal; outside of the canal, the gas pressure dropped precipitously so that high vacuum ($10^{-8}$ – $10^{-9}$ torr) was maintained throughout the rest of the instrument. Argon was chosen as the primary stripping gas because it provided a good balance of key features: (1) large molecular destruction cross section; (2) production of predominantly charge +1 ions for most elements; (3) acceptable scattering loss. A preliminary study of other gas species, such as He, Ne, N$_2$, C$_2$H$_4$ (ethylene), Kr, and SF$_6$, found that most gases, save ethylene, preferentially produce higher charge state ions[9]. The NAUTILUS does not incorporate energy-sensitive ion detectors, so analyzing charge +1 ions is crucial for limiting potential interferences. In contrast, highly charged ions are preferred for other AMS applications. All of the data presented here were collected using a mass-flow controller that does not have a positive shut-off valve. This means that even at a set-point of zero standard cubic centimeters per minute (sccm) flow, a small quantity of gas still passes through the flow controller diaphragm. This quantity of gas does not result in any appreciable scattering loss, however, it does promote charge state change for a few percent of the incident ions. Overall, this has little effect on the analysis presented here, as the normal operating condition of the NAUTILUS is with non-zero gas flow.

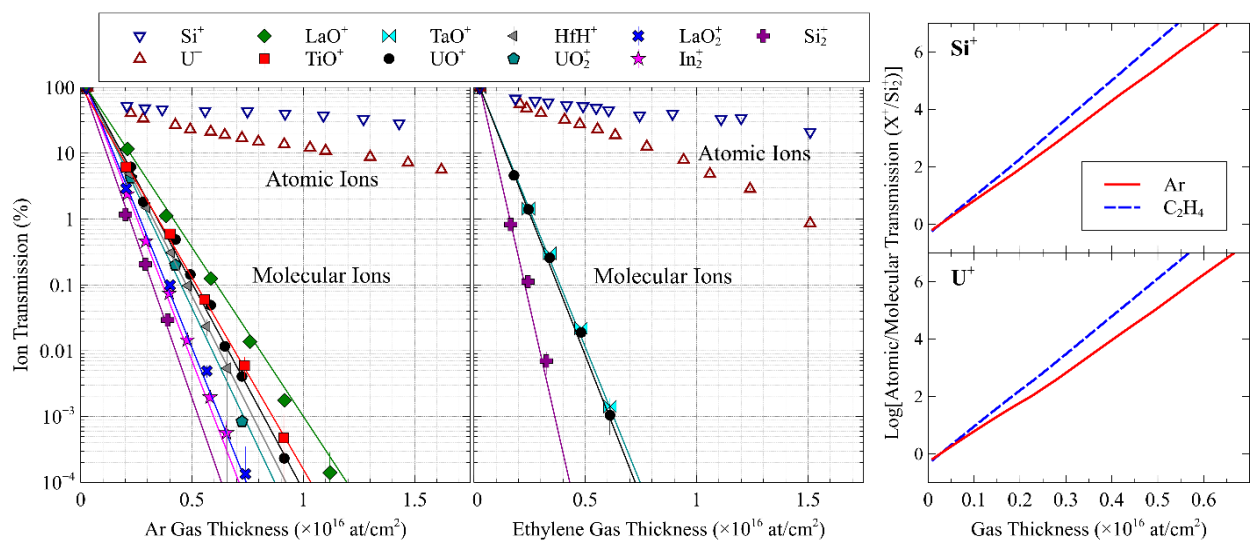

Figure 7: Comparison of molecular destruction cross sections and atomic ion transmission for Ar and ethylene (C₂H₄) stripping gases. Ethylene exhibits better transmission for atomic ions relative to its molecule destruction power

Based upon measurements and TRIM[31] calculations, which simulated scattering and energy losses in the gas cell, we determined that the majority of transmission loss is due to charge state change. Figure 6 shows a comparison of transmission and charge state change for injected Si⁺ (bottom row) and U⁺ (top row) ions through different stripping gases, including Ar, C₂H₄ (ethylene), Ne, N₂ (U only), and He (Si only). These are a small, but illustrative, example of gas species effects. As seen here and in Groopman et al.[6, 9], the projectile species also has a significant impact on transmission and charge state population. Figure 6 shows that U⁺ ions have significantly worse transmission than Si⁺, which is a function of their lower velocity and therefore larger scattering cross sections (Figure 11) in addition to a higher propensity to undergo charge state change. Charge state change includes both production of multiply charged ions and production of neutrals. We cannot measure neutralized atoms on our mass spectrometer, but can infer their populations by summing the ions we observe in higher charge states and subtracting this from the expected scattering loss based on TRIM. The inferred neutral populations are shown in Figure 6 with dashed open symbols, whereas measured ions are shown in closed symbols (charge +1: red circles; +2: blue squares; +3: green triangles). The exponential fits to TRIM-calculated transmissions did not always intercept precisely at 100% transmission, so a vertical shift in each of the inferred neutral populations may be present. Therefore, the inferred neutral populations are only qualitative.

Charge +1 ions were the dominant measureable signals for Ar and C₂H₄ stripping gases, making them good candidates for our use. Neon resulted in less scattering loss than Ar or C₂H₄, however, it produced more multiply charged ions. For Ne gas, the differences between U⁺ and Si⁺ incident ions were significant, with charge +2 becoming roughly equivalent to charge +1 for Si at gas thicknesses above 5×10¹⁵ atoms·cm⁻², while for U charge +2 briefly becomes most abundant, after which charge +3 dominates above 5×10¹⁵ atoms·cm⁻². Having different elements behave so differently to the same stripping gas is not ideal for general purpose measurements or where the behavior of each element has not been mapped. Helium stripping is often used because of its low scattering cross section, however, it predominantly produces higher charge-state ions (e.g. Vockenhuber et al.[32]) as we observed for Si and other elements we examined (not shown). Likewise, N₂ results in relatively little scattering loss compared to Ar, but produces more multiply charged ions than charge state +1. This has also been seen on the MICADAS miniature radiocarbon system[33]. The phase space of element/stripping gas combinations is extensive and the effects of different stripping gases may vary significantly for analyzed elements across the periodic table. We were, therefore, interested in identifying stripping gases that performed reasonably well for as many elements as possible. This also leaves open the option to optimize the stripping gas and transmission for a specific measurement, if necessary.

Based upon our preliminary search, Ar and C₂H₄ provide the best transmission of charge +1 ions, with C₂H₄ providing ~2× the transmission of Ar at low flow rates. Since the NAUTILUS operates as a molecule filter, atomic transmission is not the only criterion to optimize. Figure 7 shows both atomic and molecular ion transmission relative to the gas thickness of Ar (left panel) and $C_2H_4$ (middle panel). Exponential fits to the molecular transmissions are also shown. As discussed in Groopman et al.[6], the molecular dissociation cross sections appear to approximately scale with bond dissociation energy (e.g., Ta-O = 839 kJ/mol; La-O = 798; U-O = 758; Ti-O = 666; Si-Si = 310; In-In = 82[3]) though this is likely convolved with other factors such as ion velocity. Ethylene appears to be more efficient at dissociating molecules than Ar at similar gas flow rates, potentially due to its larger physical cross section. In the right panel we plot the ratio of atomic to molecular ion transmission for $Si^+$ and $U^+$ relative to $Si_2^+$ for both stripping gases. A higher ratio is better as it indicates less scattering and charge state losses for atomic ions and/or more efficient molecule destruction. For both $Si^+$ and $U^+$ at gas flow rates below $3\times10^{15}$ atoms·cm⁻², $C_2H_4$ was found to be more efficient, while Ar was more efficient above. This indicates that Ar would be the better choice for removing a large molecular background in filter mode, while $C_2H_4$ would be better for fragment mode and in cases where the molecular background is less substantial. For instance, at $3\times10^{15}$ atoms·cm⁻², molecular dissociation for Ar ranges between 1.5 and 3 orders of magnitude, depending upon the molecule, while for $C_2H_4$ it ranges between 2.5 and 3.5 orders of magnitude. Above this thickness, however, the transmission for atomic ions in $C_2H_4$ drops more precipitously than in Ar, negating the positive effects of more efficient molecular dissociation. For problems such as measuring trace heavy REEs under the intense oxide molecules from light REEs (e.g., ¹⁵⁵Gd⁺ under ¹³⁹LaO⁺ in Madagascar hibonite[6]) an Ar stripping gas maximizes the SNR. For fragment mode analyses, where only a ~2 order of magnitude reduction in molecules is necessary, $C_2H_4$ provides higher atomic transmission. For many other cases where a more modest 2 to 3 order of magnitude reduction in molecular background would suffice, $C_2H_4$ also appears to be promising.

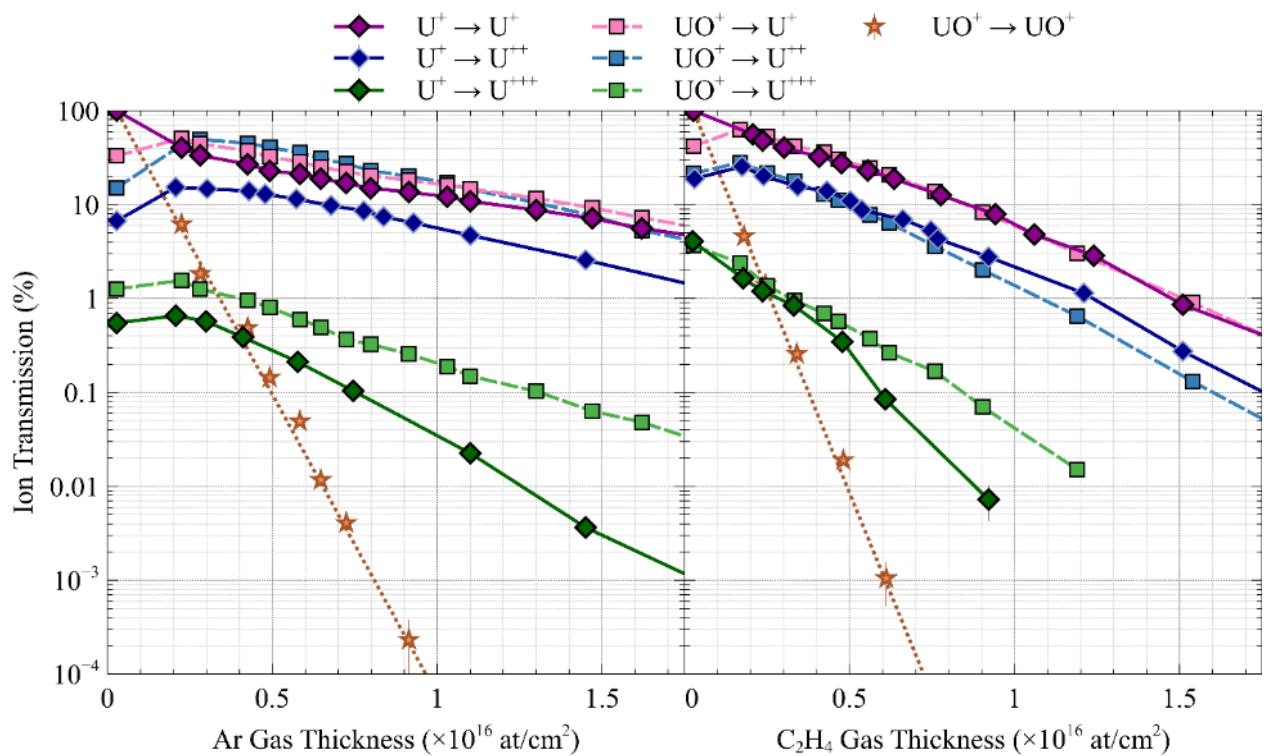

Figure 8: Comparison of the charge state distributions of U ions and fragments from injected U⁺ and UO⁺ using ethylene and Ar stripping gases. Ethylene is more efficient at dissociating UO⁺, which leads to higher transmission of U⁺ fragments for lower flow rates. Interestingly, ethylene produces similar charge state distributions for U regardless of whether the injected ions were U⁺ or UO⁺. Argon promotes relatively more U⁺⁺ from UO⁺ than from U⁺.

Figure 8 shows a comparison between Ar and $C_2H_4$ of the charge state distributions of U ions analyzed in filtering and fragment modes, from incident U⁺ and UO⁺, respectively. The transmission of UO⁺ is also shown for reference. The charge state distributions for filtering U⁺ ions are the same as in Figure 6 for both gases. The charge state distributions of U ion fragments are significantly different between the stripping gases, whereas distributions of filtered atomic U ions are similar, albeit with different scattering loss slopes. It is striking that $C_2H_4$ results in filtered and fragment charge state distributions that are nearly identical, while simultaneously having higher transmission for lower flow rates. In addition, $C_2H_4$ promotes $U^+$ ions in both measurement modes. Argon, by contrast, promotes $U^{2+}$ and $U^{3+}$ fragments from $UO^+$ more intensely than from $U^+$. While this is only one example, it is beneficial to use a stripping gas that results in the same behavior for an element, regardless of whether it is analyzed as an atomic ion or as a molecule fragment. From these data it is clear that $C_2H_4$ is the best stripping gas for fragment analysis that has been investigated so far.

**Peak Shapes**

The introduction of a stripping gas into the ion beam path results in energy losses from the ions; simulations are described in the TRIM Calculations section. From a qualitative perspective, this should result in a degradation from the ideal "flat-topped" peak shape coveted in mass spectrometry to a peak with rounded shoulders and a low-energy tail. The measured peak shapes of $U^+$ ions through Ar gas of thickness $7 \times 10^{15}$ atoms·cm$^{-2}$ for varying incident energy dispersions, e.g., SIMS energy slit widths, are shown in Figure 9. These peak shapes were measured by scanning the SSAMS EPS voltage across the 3 mm-wide slits between the SSAMS magnet and ESA. The key feature of the peak shapes is that they are asymmetric, with the high-energy (right-hand) shoulder being of lower intensity than the low-energy (left-hand) shoulder, which is followed by a lower-energy tail. This peak shape can be understood to be the convolution of the flat-topped peak produced by the slits (e.g., Figure 4) with an energy-loss distribution from the ion-gas collisions. Qualitatively, there should be fewer high-energy ions following the collision cell, where some of these have populated the lower-energy shoulder and tail, causing the low-energy shoulder to be more intense than the high-energy shoulder. The peak shape varies slightly for different gas flow rates, i.e. different magnitudes of energy loss, however, for a given gas flow rate, the peak shape does not vary with time. We, therefore, take the centers of individual peaks to be the center of mass, calculated by scanning the EPS across the full peak width or by taking a seven-point measurement across 1.5×FHWM. It is important to emphasize that because the peaks are relatively broad and their shapes do not change with time, the COM is stable throughout measurements. We therefore do not lose precision due to drift, despite the non-flat-topped peak shape. Figure 9 also shows that the incident energy dispersion of the ions from the SIMS essentially only affects the overall ion signal and does not affect the peak shape or its FWHM. In a uniform electric field, the energy dispersion of the incident ions is maintained through acceleration, so ions in the SIMS with 4.5 keV of energy with 100 eV dispersion would be accelerated in the SSAMS to 304.5 keV with 100 eV dispersion. This is true of the NAUTILUS to first order. The energy dispersion caused by ion-gas collisions in the gas cell is at least 20× larger than the SIMS energy dispersion, which is why we do not observe a significant effect from the SIMS energy slit width on the SSAMS peak shape. Since we operate the SIMS and SSAMS with relatively low MRP compared to conventional SIMS and the increased energy dispersion does not adversely impact our selectivity, we gain considerable sensitivity by operating the SIMS with a relatively wide-open energy slit, ~3× ion transmission compared to a standard 50 eV slit width. The peak shapes for other elements are similar to U and are therefore not shown.

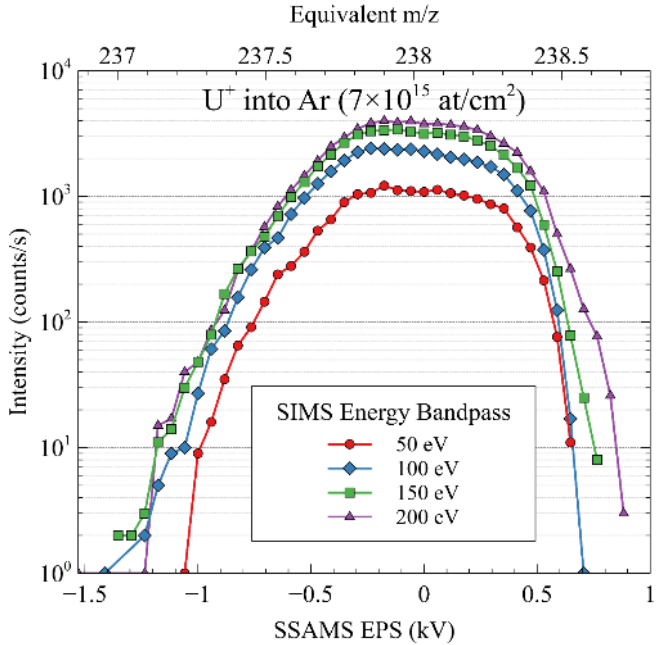

Figure 9: Comparison of peaks shapes on the SSAMS while varying SIMS energy slit bandpass. Opening the SIMS energy slit significantly increases transmission. The gas stripping cell typically results in ~2kV energy loss and ~1kV energy spread, which is far larger than the extra 50 – 100 eV energy dispersion from the SIMS. The additional energy dispersion from the SIMS is acceptable since the NAUTILUS is not run with high MRP and instead uses gas stripping to remove molecular isobars. EPS setting arbitrary though close to 0 kV for the SSAMS magnet centered on $^{238}U^+$.

## Gas Stripper Modelling

In order to convert the gas flow rate set by the mass-flow controller (sccm) into a gas thickness (atoms·cm$^{-2}$), we modelled the gas stripper based upon schematics provided by NEC. The gas thickness depends primarily on the gas flow rate, the gas stripper geometry, and the fluid flow regime, e.g., molecular, transitional, or viscous. We applied an iterative solver to the conductance equations found in Lafferty[34] to the series of concentric apertures and tubes that comprise the gas stripper geometry. This approach used the pressures read from two hot ion gauges located at different positions outside of the inner stripping canal to infer the pressure inside of the canal where the ion beam transits and collisions occur. Gas-specific correction factors were applied to the mass-flow controller and to the hot ion gauge readings. At each iteration of the calculation the Knudsen number, Kn, for each section of the gas stripper was quantified. Kn is a dimensionless quantity defined as the ratio of the mean free path length for a stripping gas atom relative to the dimensions of the gas stripper section. For low gas flow rates, the stripping gas was in a molecular flow regime, where the mean free path of the atom or molecule was comparable to or larger than the dimensions of the chamber, which we defined in the software as having Kn greater than 0.5[34]. For regions and flow rates with Kns between 0.01 and 0.5, we applied a correction for transitional flow based upon slip theory[34, 35]. These calculations were packaged into a small Python GUI, which accepts a list of flow rates and a gas species, and returns several parameters including: gas thickness with and without the transitional flow correction, in atoms·cm$^{-2}$, moles·cm$^{-2}$, or torr·cm; conductances and Kns for each section of tube; and the transition point between molecular and transitional flow. We operate well below the viscous flow regime (Kn < 0.01), so these calculations were omitted. For Ar, the transition between molecular and viscous flow occurs at 0.5 sccm or 1.75×10$^{16}$ atoms·cm$^{-2}$, while the transition for C$_2$H$_4$ occurs at 0.62 sccm or 1.84×10$^{16}$ atoms·cm$^{-2}$. A typical flow rate for filtering measurements is 0.2 – 0.3 sccm of Ar. The resulting gas thicknesses from these calculations were used to derive gas densities in the stripping canal for each flow rate, which were supplied to TRIM calculations of the scattering loss.

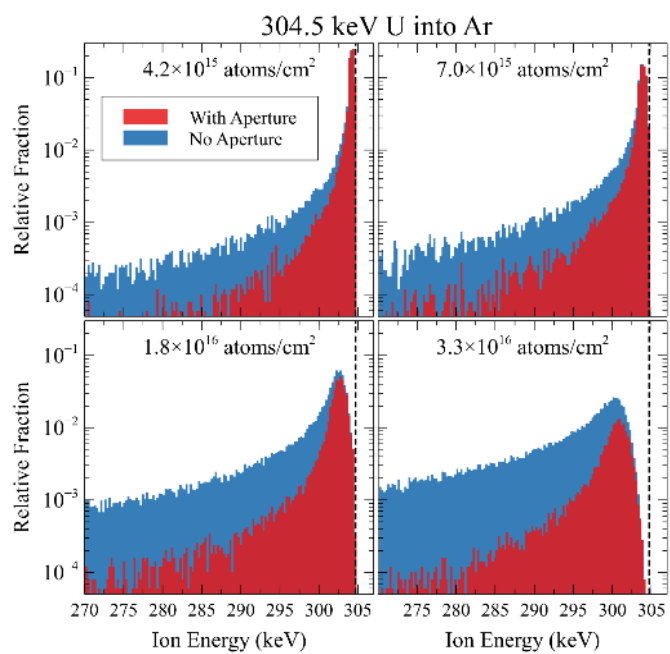

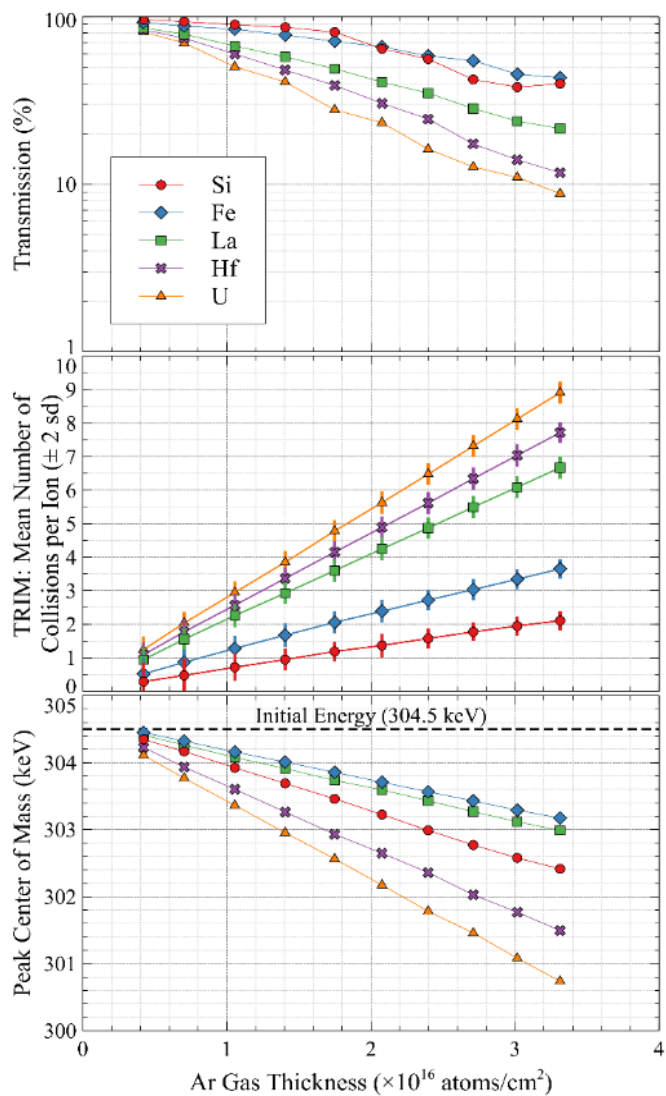

Figure 10: Simulated TRIM energy distributions of 304.5 keV U$^+$ ions into Ar. An aperture limits the acceptance angle of the SSAMS spectrometer. Typical operating conditions at 7×10$^{15}$ at/cm$^2$. Measured peak widths are 1.5 - 2 keV. See Figure 11 for integrated transmission around the peak maxima.

Figure 11: Simulated transmission and energy loss of charge +1 ions into Ar gas stripper (top panel). Transmission was calculated based upon a 2 keV window around the peak maximum. The middle panel shows the mean number of collisions per ion with 2 sd confidence bars. Molecules likely have larger interaction cross sections than atomic ions given their larger size. The bottom panel shows the energy shift of each peak relative to its initial 304.5 keV energy.

## TRIM Calculations

James Ziegler's freely available SRIM program[31] was used to simulate the transmission of various atomic ion projectiles through a range of stripping gas densities and species. Gas densities were calculated as per the description in the previous section and compound corrections were used where appropriate, e.g., for C$_2$H$_4$. The SRIM calculations provided for each ion: final position and scattering angle information, energy loss, and total number of collisions. SRIM does not simulate molecular projectiles, nor does it calculate charge state change for the ions. We therefore used the SRIM calculations primarily to measure scatting and energy loss for atomic ions. The limit for an ion beam injected by the SIMS to be measured on the SSAMS EM is less than 1 pA, meaning that on the order of 10$^6$ ions/s transit the gas stripper canal. For $^{238}$U$^+$ ions, 304.5 keV corresponds to a velocity of nearly 5×10$^7$ cm/s, yielding a transit time through the gas stripper canal of approximately 1 μs. Therefore, on average, only one ion would be present in the entire stripping canal at a time and any effects from other ion-gas collisions or ion beam repulsion can be safely ignored; SRIM's serial simulation of individual ions transiting the gas is, therefore, appropriate.

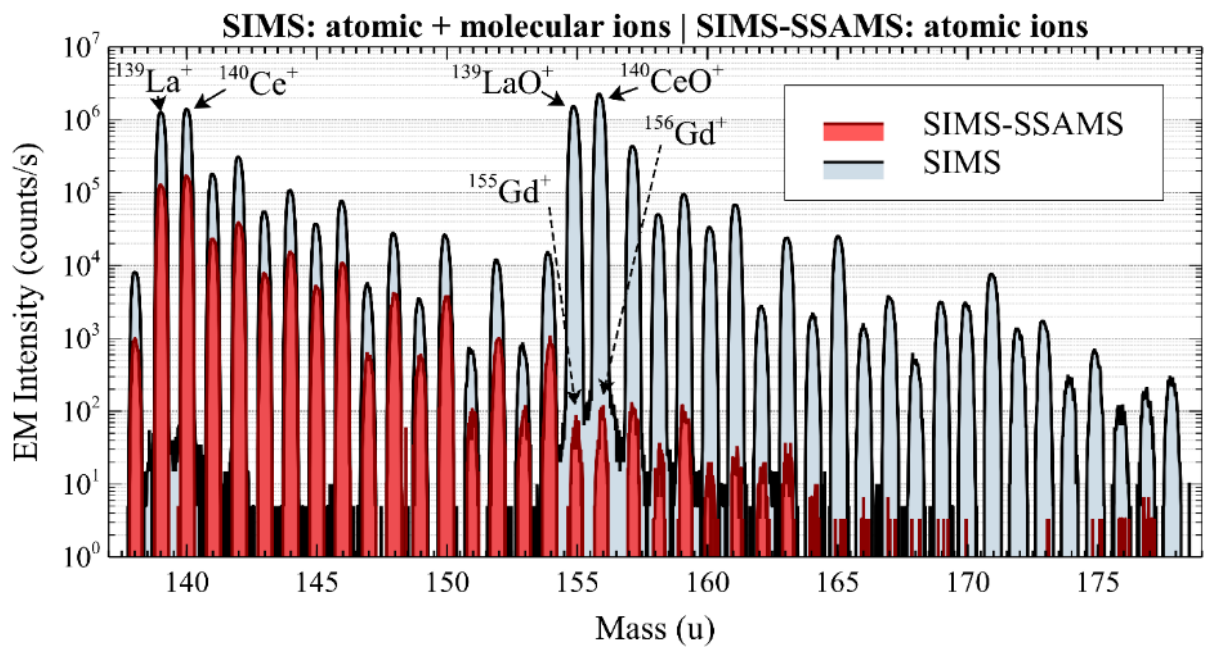

Figure 12: Mass scan comparison over the REEs from a Madagascar hibonite measured on the SIMS (blue) and SSAMS (red) EM detectors of the NAUTILUS. The SSAMS enables direct measurement of trace isotopes only $10^{-3}$–$10^{-4}\times$ as intense as the molecular background. Figure reproduced from [6] with permission from the author/copyright holder.

Figure 10 shows the simulated energy distributions of monochromatic 304.5 keV U ions transiting Ar gas for four different gas thicknesses. An aperture before the SSAMS magnetic sector limits the angular acceptance of ions into the mass spectrometer to 13.5 mrad half-angle from the center of the gas stripping canal, whose effect is shown in the plot. The majority of excluded, high-angle ions are also those with the largest energy loss. The peak shapes show good agreement with those measured on the NAUTILUS (Figure 9). This is despite the TRIM calculations employing a monochromatic incident ion beam whereas the true incident SIMS beam has an energy dispersion of 150 eV. The energy dispersion effects from the gas stripper are the dominant factor affecting peak shape on the SSAMS. Based upon Figure 9, the full peak width at 1% peak height is approximately 2 keV wide. Ion transmission was calculated by integrating this 2 keV window around the center of mass (COM) of each apertured TRIM energy distribution. The energy-focusing effects of the SSAMS ESA are not included.

The resulting ion transmissions, average number of collisions per ion, and mean energy losses are shown in Figure 11 for the elements: Si, Fe, La, Hf, and U with Ar stripping gas. The transmissions of each element, save Si, follow clear exponential trends with increasing gas pressure. The transmission behavior of Si was reproducible in SRIM, however, and we cannot provide an explanation, other than noting that it was the only element tested of lower mass than the target atoms. From Gryziński[36] it is known that the Coulomb interaction cross section varies inversely with the velocity of the ion, where all ions in the NAUTILUS have much larger velocities than the average velocity of the room-temperature stripping gas atoms. Therefore, to first order, the interaction cross section of a slower $U^+$ ion will be larger than a $La^+$ ion, etc. This effect is borne out in the middle panel of the TRIM calculations (Figure 11), which shows the average number of collision events per ion for each element. For instance, at a gas thickness of $1\times10^{16}$ atoms·cm⁻², $U^+$ ions experience an average of 3 collisions in the gas cell to only 1 for $Fe^+$ ions. The number of collisions and, therefore, inelastic energy loss events yields a larger average energy loss for heavier ions than light ones (Figure 11, bottom panel). For typical Ar flow rates used in filtering and fragment modes ($<1\times10^{16}$ atoms·cm⁻²), the average energy shift of the peak is between 0.5 and 1 keV. The number of collisions per ion also plays a role in determining the final average charge state abundances, though a full discussion of such effects is beyond the scope of this paper. However, it is clear that the electronic structure of each element also plays a pivotal role in determining the equilibrium charge state populations. Figure 1 in Groopman et al.[6] shows the transmission of $^{139}La^+$ and $^{180}Hf^+$ in Ar, however, the relative transmission of La is considerably less than predicted by SRIM based upon the somewhat small velocity difference alone, indicating losses into higher charge state channels are more problematic for La than for Hf.

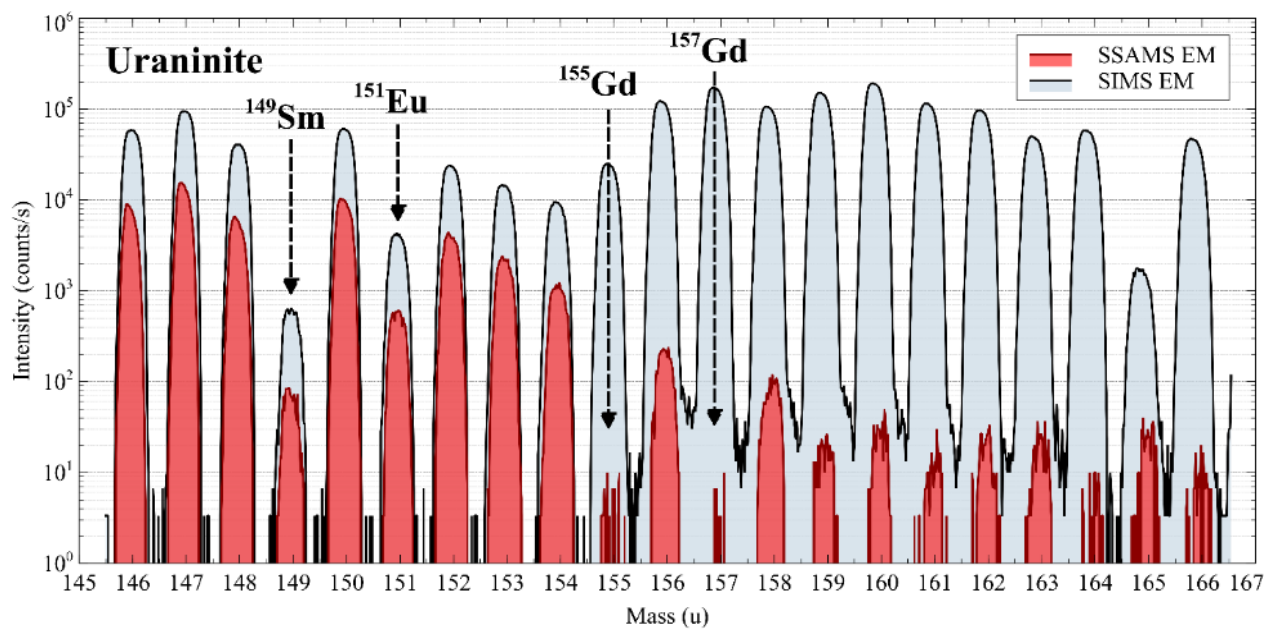

Figure 13: Mass scan comparison over the mid-to-heavy REEs measured on the SIMS (blue) and SSAMS (red) EM detectors of the NAUTILUS. The SSAMS mass scan illustrates the removal of the intense molecular background visible in the SIMS, allowing for direct measurement of fission products and n-capture depletions. These SIMS spectrum cannot be deconvolved using conventional energy filtering methods because the isotopic abundances are non-terrestrial. Figure reproduced from [5] with permission from the author/copyright holder.

**Comparison to Standalone SIMS and AMS Techniques**

The NAUTILUS features a combination of two mature analytical techniques, SIMS and AMS, so readers may be inclined to draw direct comparisons to each standalone system. The combination of instruments, however, provides a new paradigm for certain analyses. The analyses where the NAUTILUS excels overlap the capabilities of each standalone technique, but they do not encompass the full range, so readers must be careful when drawing comparisons. We argue that the NAUTILUS is more SIMS-like than AMS-like, which is predominantly a function of the NAUTILUS interacting with samples as a SIMS instrument does and the SSAMS being treated as a specialized molecule filtering "detector". However the analyses where the NAUTILUS excels occupy the middle ground, where the drawbacks of one technique are complemented by the strengths of the other. For instance, SIMS is challenged by the presence of molecular isobars, especially at high mass, and often requires inferential corrections to be made to actinide and REE measurements when MRP requirements are too high, e.g., hydride interferences for the former[14], or when the molecular background is too intense or complex, as for the latter[17, 18]. These corrections rely on careful calibration, often the assumption of non-perturbed isotopics, and a lack of certain nuclear isobars. By removing molecules and enabling direct measurement of isotopes, AMS complements SIMS. SIMS provides μm-scale spatial resolution for imaging and analysis of small features, which are often not the target of AMS analyses. Furthermore, the analytical dynamic range and precision of the NAUTILUS is to first order, more similar to a SIMS instrument than to an AMS, which is predominantly a function of the quantity of the sample consumed.

The majority of AMS analyses involve ultra-trace isotope measurements. For instance, $^{14}C$ exists in one part in $10^{12}$ in the atmosphere and many radiocarbon analyses require measurement of $10^{-15}$ abundances. This dynamic range is vastly larger than the roughly 9-10 orders of magnitude available on the NAUTILUS when combining FC and EM detectors and a SIMS front-end, though we often limit analyses to the dynamic range of an EM to use less-intense primary ion probes with higher spatial resolution. Although the gas stripping process is stochastic, where there exists some probability that molecules will make it through the gas cell intact, it remains relatively simple to reduce the molecular background by 5-7 orders of magnitude without significantly compromising atomic ion transmission. Therefore, molecular signals are consistently below the statistical limits of our analyses and their transmission can be easily modulated by adjusting the gas flow rate into the stripping cell. This level of molecule reduction would clearly be inadequate for radiocarbon analyses, but these analyses would already be atom-limited given our focus on micron-sized features of interest and the injectable ion signal from a SIMS instrument. Other typical AMS measurements include $^{10}$Be, $^{36}$Cl, and $^{26}$Al, where abundances range from $10^{-12}$ to $10^{-18}$. This is the crux of the difference between the NAUTILUS and other AMS techniques and should be emphasized. The qualification of whether a filtered signal is "molecule-free" has significantly different meaning when we measuring single ions from aA to pA ion beams (1-$10^6$ ions·s$^{-1}$) injected from the SIMS versus a conventional AMS instrument measuring single ions from μA (>$10^{12}$ ions·s$^{-1}$) or more intense beams. In our NAUTILUS analyses we have yet to observe any molecular or multiply charged isobar interferences that cannot be adequately addressed by modulating the flow rate in the gas cell, which lowers the interfering species' signal to below statistical precision. In addition, the sensitivity benefits of measuring certain elements in fragment are further bolstered by a significant reduction in the likelihood of potential interferents since the molecule signal is being converted into the signal of interest, with energy partitioning.

To expand upon this point, it is important to emphasize several aspects of SIMS and SSAMS that benefit the NAUTILUS. It is important that we analyze only charge +1 ions with the SSAMS, because this minimizes the potential number of multiply charged interfering species, especially for high-mass elements, since there are no lower charge states. In addition, the sputtering process in SIMS is exceptionally poor at producing multiply charged atomic and molecular ions for elements heavier than Al, Si, and Ca[37]. SIMS mass spectra only rarely exhibit peaks at fractional nominal masses, indicative of the propensity for producing singly charged ions. If complex interfering multiply charged ions are suspected to be present, fragment mode analyses can elucidate their composition. The charge state distributions of the fragments can also be measured to verify their identities. So far we have only observed full molecule dissociation on the SSAMS and have not observed complex molecules fragmenting into, e.g., LaO$^+$ and O$^+$ from LaO$_2^+$, where LaO$^+$ could subsequently interfere with Gd$^+$ from GdO$^+$ in fragment mode analyses.

Standalone SIMS, including large-geometry (LG) instruments such as the Cameca ims 1280 or the Australian Scientific Instruments sensitive high-resolution ion microprobe (SHRIMP), remain complementary to the NAUTILUS. Gas stripping on the NAUTILUS is analogous to MRP on SIMS. The NAUTILUS excels at analyses of high-mass elements where molecular isobars are increasingly difficult to separate via MRP alone and where EPS provides the widest range of measureable isotopes for a given magnetic field setting. Additionally, the transmission of ions is greater in the NAUTILUS using gas stripping (in filtering mode and more so in fragment mode) than for other SIMS techniques using energy filtering for elements such as the rare earths. Transmissions of REEs were found to be 1 – 10% for Cameca small and large geometry instruments using energy filtering and/or high MRP [17, 18, 38-40], ~20% for SHRIMP-RG using high MRP and moderate energy filtering [41], and 1 – 10% for NanoSIMS using energy filtering[42], compared to 10 – 50% transmission on the NAUTILUS. Transmission using energy filtering is dependent upon the element of interest, as each element has a unique secondary ion energy distribution. Nominal transmission is not the only metric that governs instrumental sensitivity, however. The composition of the matrix is vitally important to any SIMS measurement. For certain trace elements, such as Gd in Madagascar hibonite (Figure 12)[6] (6 μmol·mol⁻¹), whose ion intensity is 4 orders of magnitude less intense than the oxide molecular background (La and Ce each 0.4 atom %), high mass resolving power and energy filtering will be unable to resolve the interferences, whereas this measurement is fairly routine with the NAUTILUS. The requisite amount of energy filtering and/or MRP also depends upon the matrix and will affect overall transmission. For matrices where the isotope of interest has comparable or greater intensity than the interfering molecule, SHRIMP-RG or other LG-SIMS could perform the measurement with high MRP and minimal transmission loss. LG-SIMS also outperforms the NAUTILUS for measurements of low-mass elements, e.g., O, where adequate MRP is easily achieved with minimal transmission loss.

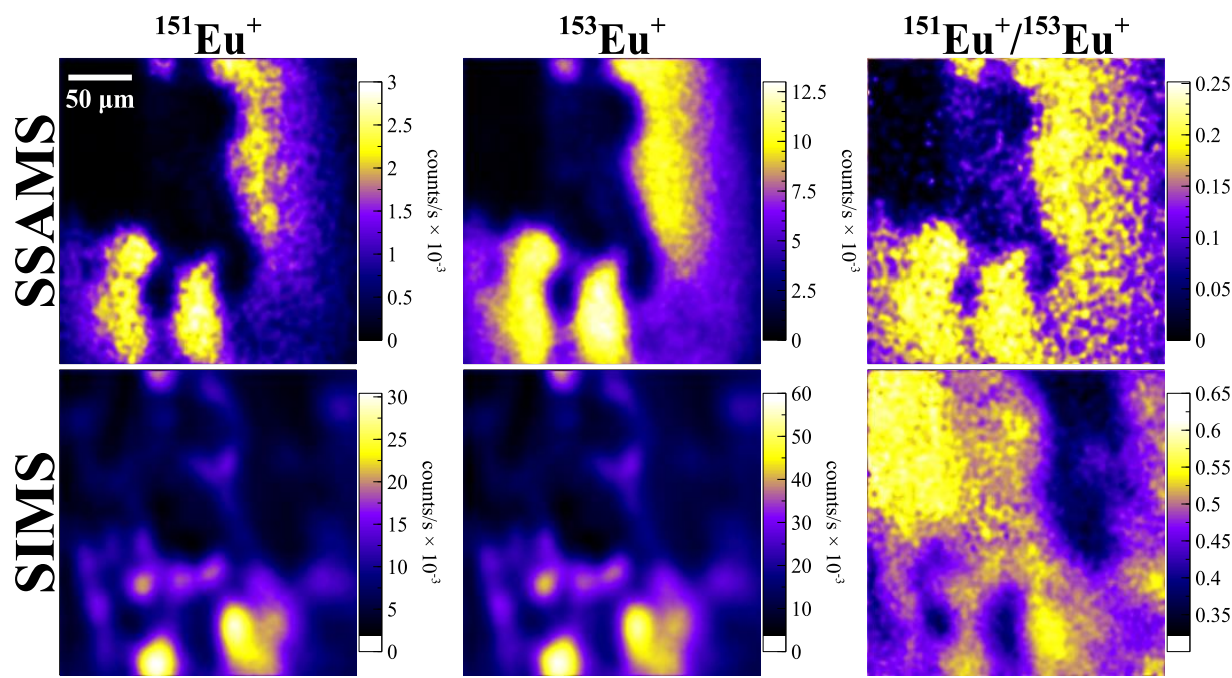

Figure 14: Europium isotope ratio imaging on the SSAMS (top, molecule-free) and SIMS (bottom) EMs of the NAUTILUS. Europium is concentrated in uraninite and not in aluminous phosphate. Europium-151 is highly depleted due to n-capture (more than [153]Eu), but this is obscured by a large and complex molecular background on the SIMS. The SIMS images contain any false-positive features and the isotope ratio image is in fact nearly the inverse of the true ratio image ([151]Eu/[153]Eu$_{terrestrial}$ = 0.916). Figure reproduced from [5] with permission from the author/copyright holder.

Figure 13 and Figure 14 further demonstrate the comparison between traditional SIMS and the NAUTILUS for analysis of REEs in isotopically perturbed and heterogeneous nuclear material from the Oklo natural nuclear reactor. Figure 13 shows a comparison of mass scans across the mid-to-heavy REEs on the SIMS alone and on the SIMS-SSAMS. Several isotopes with large n-capture cross sections are indicated with arrows, e.g., [149]Sm, [151]Eu, [155]Gd, and [157]Gd. All of these isotopes are significantly depleted due to the reactor operation, with the isotopes of Gd being depleted to near-zero. All of these isotopes are important n-fluence indicators, however they are also obscured by a large molecular background in conventional SIMS. In the cases of Gd and Dy isotopes, their abundance is so low relative to the background that they may be unresolvable even through energy filtering and high MRP. Additionally, REE measurement schemes that rely upon deconvolution of REEs from oxide molecules cannot work in perturbed isotopic systems, since the deconvolution requires the assumption of terrestrial isotopic composition. We can, however, make precise and accurate isotope ratio measurements relatively easily on the NAUTILUS using molecule filtering or fragment analysis. The problem of perturbed isotopics hampering deconvolution is further compounded by the heterogeneous matrix of these nuclear samples. Figure 14 demonstrates isotope ratio images taken at masses 151 and 153, isotopes of Eu, on both the SIMS and SSAMS EMs. Europium is concentrated in the uraninite and depleted in the adjacent aluminous phosphate. The direct SSAMS images clearly show this delineation and allow for an accurate isotope ratio image to be calculated. This shows strong depletions in the $^{151}$Eu/$^{153}$Eu ratio in the uraninite (terrestrial $^{151}$Eu/$^{153}$Eu = 0.916). By contrast, the SIMS images contain many regions of high ion intensity and structure that do not correlate to the true Eu signal. Perversely, the $^{151}$Eu/$^{153}$Eu ratio image on the SIMS is nearly the inverse of the direct image on the SSAMS due to more intense molecular interferences from the aluminum phosphate. Deconvolution corrections require the assumption of a homogeneous matrix, or knowledge *a priori* of the matrix composition across the sample using comparable standards, which is clearly not the case here. Like the spectra in Figure 13, the molecular ions from both matrices are more intense than the atomic ions, making these difficult analyses for conventional SIMS.

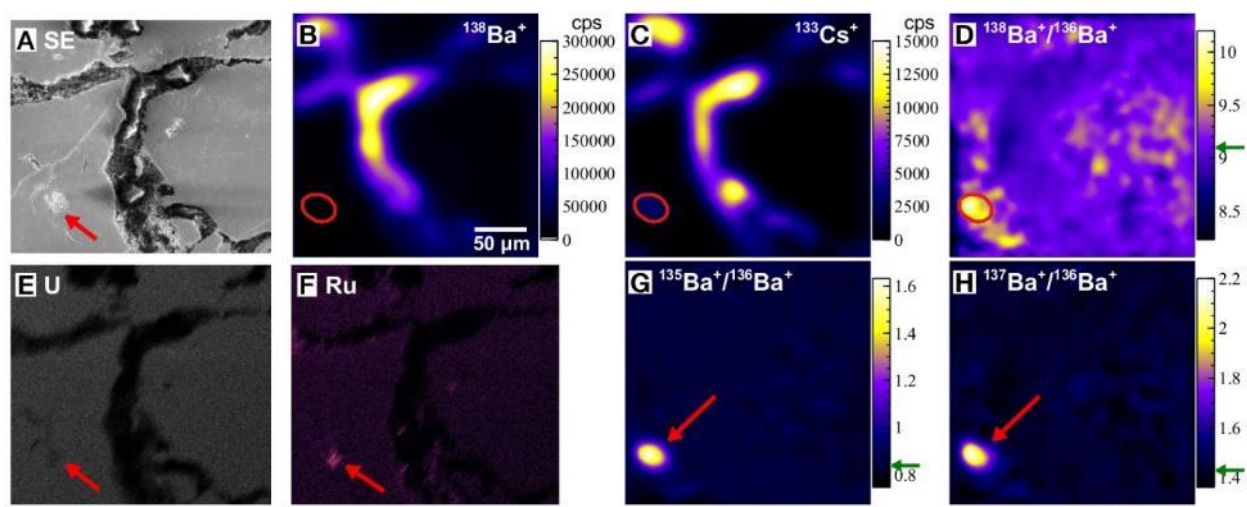

Figure 15: Molecule-free isotope ratio imaging identified fissionogenic $^{135,137}$Ba in a localized spot within a uraninite grain. Ba Cs. SEM-EDX found this spot to contain Ru metal and sulfides. The Ru phases formed ~4 years after reactor startup and captured $^{135,137}$Cs in similar abundance. The majority of Ba and Cs is concentrated in aluminous phosphate, though it has terrestrial isotopic composition (indicated by arrows in panels d,g,h). SE and SIMS images shifted horizontally relative to each other by ~50 microns. Figure reproduced from [5] with permission from the author/copyright holder.

SIMS instruments such as the Cameca ims 1280 and NanoSIMS 50(L) also incorporate multi-collection capabilities, which benefit certain analyses. The NAUTILUS is a single-collector instrument given its MS-MS design, though it is able to interleave measurements on its different detectors. There are several tradeoffs between multi-collector and single-collector instruments. For instance, the benefits of single-collector SIMS instruments are that an arbitrary number of isotopes may be measured serially in a given analysis and that inter-detector calibration is not necessary. However the duty cycle for each species is inversely proportional to the total number of analytes. Multi-collector instruments provide a higher measurement duty cycle due to parallelism and isotope ratios do not require time interpolation, which is important for rapidly changing signals, e.g., particle analyses. The downsides of multicollection are that number of analytes are limited by the number of detectors and that the detectors, such as EMs, must be inter-calibrated and corrected for ageing. While magnet hopping during multicollection is occasionally useful, only specific detector trolley positions and schemes work. The NAUTILUS is operated as a single-collector instrument where the SIMS frontend may be cycled through an arbitrary number of masses, however, due to the relatively slow switching speed of the SSAMS magnet, this field is usually fixed for a given analysis. This limits the masses selectable on the SSAMS to ±6.5% of the magnet's central mass.

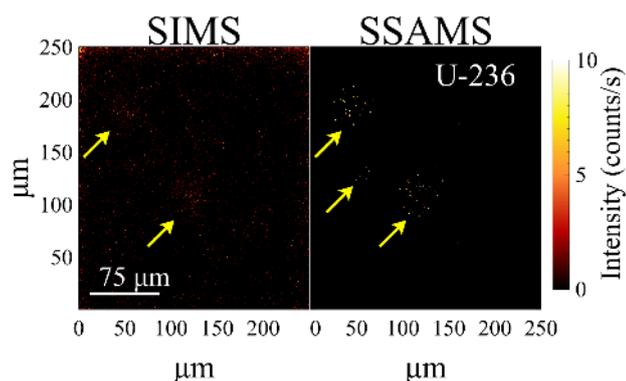

Figure 16. Comparison of SIMS and SSAMS ion images as m/z 236 of CRM U500 particles in a field of monazite "dirt". Only two large particle clusters are clearly visible on the SIMS-only image, while a third smaller one is easily seen in the SSAMS image. Figure reproduced and modified from [4] with permission from the author/copyright holder.

For measurements where large jumps are required, such as referencing abundances of high-mass minor elements to a low-mass major elements, the SIMS magnet is switched and the major element is measured there, while the high-mass ions are measured on the SSAMS. This requires inter-calibration of the SIMS and SSAMS EMs and FCs using well-known standards. Being a single-collector instrument also means that considerably less time is required to set up and change a given analysis than is required on a multi-collector. In addition, many single collector analyses spend the majority of their time counting the least abundant isotopes with less time spent on the major isotopes; therefore multi-collection often does not necessarily provide a large increase in precision. Another practical benefit of the NAUTILUS using low MRP is that tuning parameters change negligibly day-to-day and most analyses can be set up in well under an hour following startup. The low MRP also results in high instrument stability during long analyses. Most instruments are compared based upon their nominal capabilities instead of their practical ones, e.g., the 50 nm spatial resolution of a NanoSIMS or the >20,000 MRP of an ims 1280. These extreme figures are almost never used in practice because of the lack of sufficient ion signal or the presence of instrument instabilities that prevent precise isotope ratio measurements, or the increased tuning time required to achieve them. Instrument setup time and throughput are important practical characteristics that affect the science output of a lab, though the latter is often only discussed as far as automating the repetition of a single analysis, such as particle searching or automated sample exchange. It has been our attempt to discuss both the nominal and practical aspects of the NAUTILUS, with an emphasis on the latter, including benefits and drawbacks of its design. This we believe is the fairest way to make a qualitative and quantitative comparison between other SIMS and AMS instruments, which are complementary to the NAUTILUS and vice versa.

## Novel Capabilities

The combination of SIMS and SSAMS in the NAUTILUS enables several novel capabilities, including raster ion imaging without molecules and dual EPS. The geometries of the SIMS and SSAMS instruments individually are fixed as double-focusing mass spectrometers, however the length of the coupling section between them has no special significance and can be varied, with extra length being compensated for by additional deflectors and lenses. This opens the possibility of adding new ion sources for injection directly into the SSAMS via an electrostatic switch, similar to the ETH TANDEM[19], or inserting new "filters" such as a neutralization and resonance re-ionization scheme for removing nuclear isobars.

### Direct, Molecule-Free Raster Ion Imaging

Direct, molecule-free raster ion imaging on the NAUTILUS represents its most significant analytical capability. Single-spot microanalysis free from molecular isobars is a unique capability among SIMS instruments, however imaging with this capability gives us leverage over an array of relatively intractable problems for SIMS. As discussed previously, Figure 14 shows a comparison of direct, molecule-free imaging on the SIMS-SSAMS versus imaging on the SIMS alone, which was full of interfering species from the heterogeneous matrix. Figure 15 shows an additional example of where direct isotope ratio imaging of Ba and Cs isotopes allowed us to locate an anomalous hotspot containing fissionogenic $^{135}$Ba, $^{137}$Ba, and $^{138}$Ba that was otherwise indistinguishable via SIMS[5]. This hotspot was later correlated to the presence of Ru metal and sulfides, which apparently captured live fissionogenic $^{135}$Cs and $^{137}$Cs within 5 years of the Oklo natural nuclear reactor ceasing criticality. The powerful comparison between SIMS and SSAMS in Figure 14 and the scientific discovery shown in Figure 15 illustrate the importance of direct ion imaging for rapidly locating regions of interest in complex samples. Direct, uncorrected ion imaging eliminates the uncertainty regarding whether potential features of interest are simply the result of varying topography and/or molecular background. Other examples of the power of direct, uncorrected ion imaging are given in Willingham et al.[4] and Figure 16. The NAUTILUS was used to collect rapid isotope images on m/z = 236, where $^{235}$U$^1$H$^+$ typically interferes with $^{236}$U$^+$, and remains unresolvable with high MRP. Instead of requiring a correction based upon the inferred hydride abundance through measuring $^{238}$U$^1$H, which does not work if Pu is present in the sample, we were able to identify small particles containing $^{236}$U directly. In this particular example a background of monazite "dirt" was placed over NIST CRM U500 particles on a vitreous carbon planchet. Only 2 particles are clearly distinguishable at mass 236 on the SIMS-only image, while a third, much smaller particle is visible on the SSAMS. This type of rapid screening for an isotope of interest does not rely on the measurement of major isotopes to identify potential candidates for spot analysis nor does it require measuring $^{238}$U$^1$H to perform an inferred correction.

With the SSAMS on the NAUTILUS treated as a large molecule-filtering detector, raster ion imaging was achieved much as it is on other SIMS instruments. As the primary ion beam is rastered across the sample, a set of deflector plates following the SIMS immersion lens (the dynamic transfer optical system (DTOS[10]) is energized to deflect the secondary ions coaxially through the mass spectrometer. This system corrects for the trajectories of ions produced off of the immersion lens axis by the rastered primary beam, and is especially import when the primary raster size is larger than the ion microscope's static field of view. Once the ions from the rastered beam are sent coaxially through the mass spectrometer, we deflect them into the SIMS detectors or inject them into the SSAMS in the standard fashion. The two pulse trains controlling the timing of the primary and secondary raster patterns may be delayed as necessary; time-of-flight delays are also calculated on a mass-by-mass basis and applied to the gate delay on the EM counters. The NAUTILUS is not the first AMS system to employ imaging. For instance, Freeman et al.[43] describe their use of a radiocarbon accelerator and a sputter source to capture SIMS images of $^{12}$C and $^{13}$C. The key distinction is that this was not the same source used for radiocarbon measurements and the sources were mutually exclusive. By contrast, molecule-free SIMS spot analyses with the NAUTILUS are essentially identical to imaging analyses except that the primary and secondary beams are rastered instead of kept static.

**Dual Electrostatic Peak Switching**

Electrostatic peak switching on both SIMS and SSAMS simplifies synchronization of the two magnets in the NAUTILUS' MS-MS configuration. As described in Groopman et al.[6], the duty cycle of the instrument is improved due to more rapid switching times across the ±6.5% EPS range. A significant benefit is that masses may be analyzed in any order for a given magnetic field setting since EPS does not affect magnet hysteresis. Therefore, for analyses such as U-Th-Pb dating, when the magnet is tuned near the Pb isotopes, $^{206}$Pb$^+$ or $^{208}$Pb$^+$ may be measured and centered upon first before measuring the usually underabundant $^{204}$Pb$^+$ peak. The magnet flight tube, when biased, acts as a lens, with different focal properties for accel-decel and decel-accel modes[6]. The SIMS spectrometer lens was calibrated to compensate for this effect automatically through the use of a gain factor when switching the EPS. The image of the beam is, therefore, nearly uniform for any EPS bias as it transits the mass spectrometer. This effect is far less pronounces on the SSAMS, so there is currently no correction applied. These effects can lead to instrumental mass fractionation, which must be accounted for through the use of standards, as is typical with SIMS or any mass spectrometry technique. Dual EPS can also be used to collect molecule-free mass spectra over the ±6.5% EPS range, as in Figure 12 and Figure 13. For larger ranges, mass scans at different magnetic fields can be stitched together.

**Ultra-low Measurement Background**

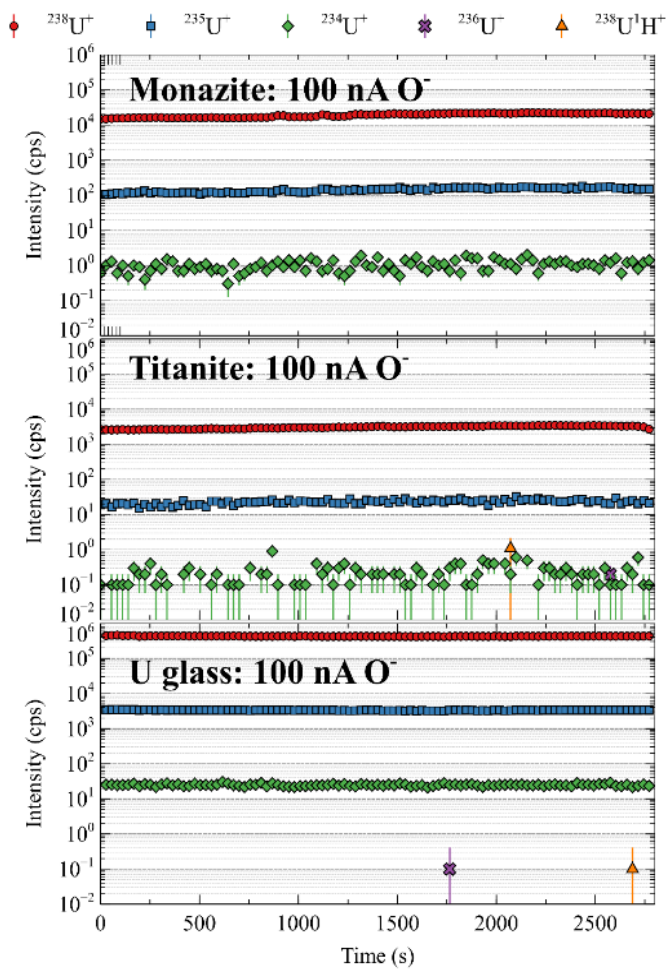

Figure 17: Depth profiles from monazite (NMNH# R14013), titanite (Grenville skarn[1]), and U Canary glass with natural U isotope composition showing extremely low background/noise. On each cycle, count times in seconds for $^{238}U^+$, $^{235}U^+$, $^{234}U^+$, $^{236}U^+$, and $^{238}U^1H^+$ were: 1, 2, 5, 10, 10, for a total of 1000 s each for $^{236}U^+$ and $^{238}U^1H^+$ in each profile. Uranium hydride abundances are typically ~$10^{-3}\times$ the intensity of the adjacent peak in SIMS, so we would expect $^{238}U^1H^+$ count rates of 20, 3, and 400 cps for the monazite, titanite, and U glass, respectively, and $^{235}U^1H^+$ count rates of 0.1, 0.01, and 3 cps on the SIMS alone. The gas stripping efficiently removes these molecules, with only two noisy cycles each present in the titanite and U glass profiles. With a natural abundance of <$10^{-11}$, no counts of $^{236}U^+$ are expected, as seen in the monazite profile. These data were collected prior to adding TVS diodes to the motor-generators which further decreased detector noise.

One of the most significant features of the NAUTILUS is its ultra-low background relative to commercial SIMS instruments. Several factors influence this, including the design of the EM electronics, high impact energy of ions on the EM first dynode, filtering by the gas cell and the MS-MS design, and the Faraday cage around the accelerator, despite being nominally for electrical safety. The sensitivity of a measurement is ultimately limited by the achievable SNR. For trace isotope and element analyses, where signals are small, instrument background and detector noise provide a floor for measurement accuracy, while Poisson counting statistics provide a floor for measurement precision. When signals and noise are of comparable magnitude and cannot be deconvolved, the noise perversely adds meaningless statistical precision to the inaccurate measurement.

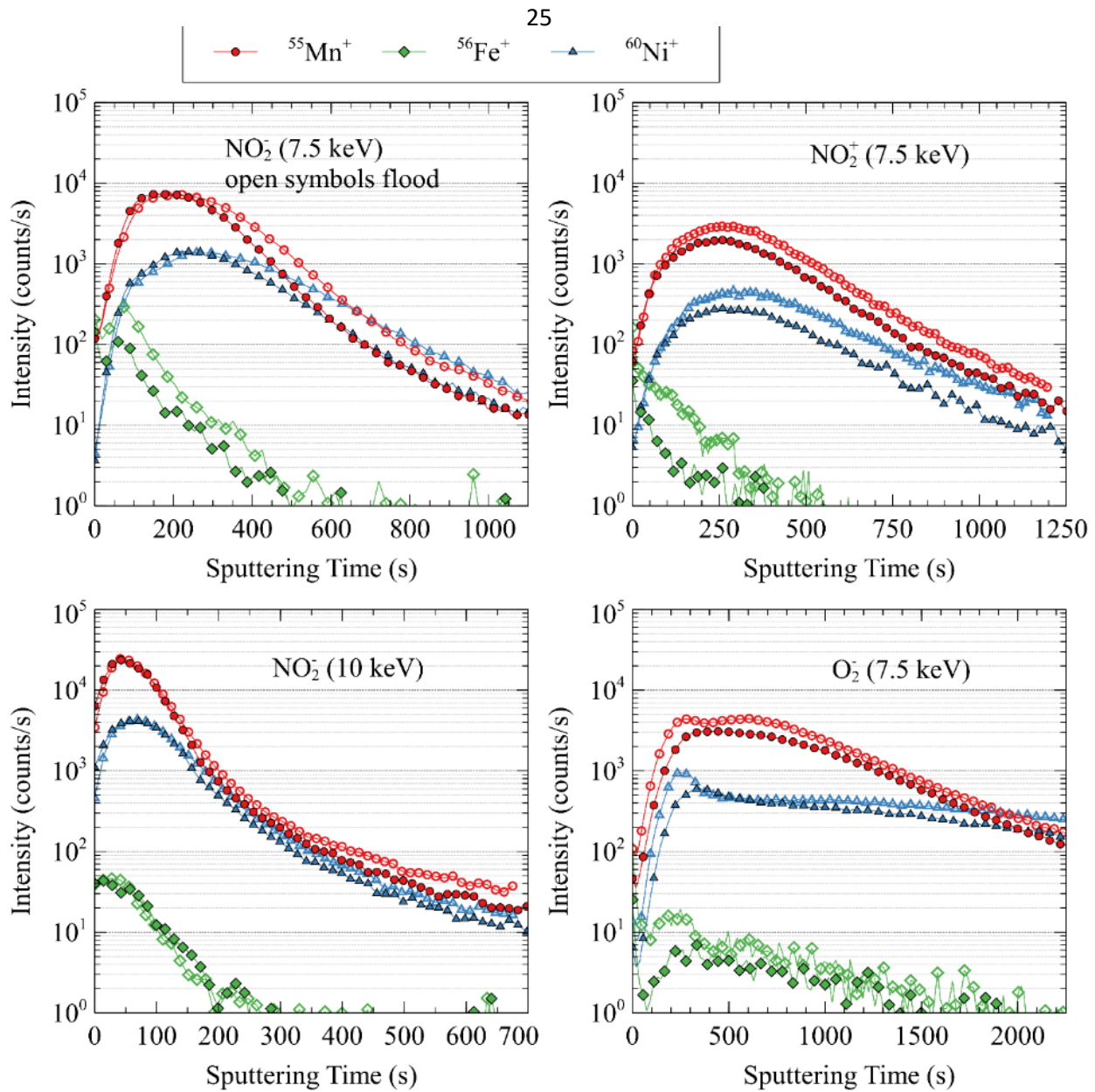

Figure 18: Comparison of using an $NO_2^{\pm}$ ion beam under different conditions to depth profile a $^{55}$Mn (55 keV, peak ~50 nm deep) and $^{60}$Ni (60 keV) ion implant in Si metal. Some Fe was co-implanted and/or due to surface contamination. Closed symbols are depth profiles without any flooding; open symbols include sample flooding with $CF_4$. Markers are thinned for visual clarity. Secondary ion energy is 4.5 keV, so impact energies are, e.g., 3 keV for $NO_2^+$ (7.5 keV) and 14.5 keV for $NO_2^-$ (10 keV). $NO_2$ exhibits better ion yields, depth resolutions, and sputtering rates than $O_2$. $CF_4$ tends to enhance ion yields when they are low, e.g., $NO_2^+$ and $O_2^-$, but only appears to affect the sputter rate when yields are higher. This sample was provided by Amy Jurewicz and Donald Burnett.

Figure 17 shows three U isotope depth profiles of monazite (National Museum of natural History (NMNH) # R14013, India), titanite (Grenville skarn[1], Canada), and U (Canary) glass, with a range of U abundances. All profiles were measured in early 2018 prior to subsequent noise-reducing modifications described later in this section. Each profile was made under primary ion beam intensities of 100 nA of O$^-$, simulating conditions where one would expect the largest abundance of interfering species. Larger ion probe currents are not typically used for SIMS analyses. On each cycle, count times for $^{238}U^+$, $^{235}U^+$, $^{234}U^+$, $^{236}U^+$, and $^{238}U^1H^+$ were: 1, 2, 5, 10, and 10 s, for a total of 1000 s each for $^{236}U^+$ and $^{238}U^1H^+$ in each profile. Uranium-hydride abundances are often ~$10^{-3}\times$ the intensity of the adjacent peak in SIMS, so we would expect $^{238}U^1H^+$ count rates of 20, 3, and 400 cps for the monazite, titanite, and U glass, respectively, and $^{235}U^1H^+$ count rates of 0.1, 0.01, and 3 counts/s (cps) on the SIMS alone without energy filtering. The gas stripping efficiently removes these molecules, with only two noisy cycles present in each of the titanite and U glass profiles. With a natural abundance of $<10^{-11}$ relative to $^{238}U$, no counts of $^{236}U^+$ are expected, as seen in the monazite profile. From these three measurements, a combined noise and instrument background of $1.6\times10^{-3}$ cps is observed, with the majority of these counts coming from a single cycle in the titanite measurement ($3.5\times10^{-4}$ cps without this cycle). For comparison, the detector dark noise alone on Cameca ims 1280 EMs has been reported to be between $2.4\times10^{-4}$ - $2\times10^{-3}$ cps (measuring at mass 5). ETP lists maximum dark current for 14133H EM to be $5.5\times10^{-3}$ counts/s. Instrumental background including tails from adjacent masses are typically higher than dark current alone. Instead of measuring a truly blank low-mass "isotope", we demonstrate our measurement of signals where populous adjacent-mass isotopes would be expected to interfere.

During early 2018, we identified and remedied several sources of electronic noise on the SSAMS, which were yielding intermittent counts on our EM detector. We discovered that the SSAMS magnet flight tube, when biased by more than 10 kV was capacitively coupling to the magnet Hall probe introducing intermittent ringing into the system, which was detectable on the EM. Placing the Hall probe inside of a 2"×5" block of PTFE reduced the coupling and eliminated the noise. Intermittent noise under high load conditions (e.g., U analysis, ~8 kW to the SSAMS magnet alone) was removed by adding bi-directional 190V TVS diodes between each of the three phase legs of the motor-generator output and neutral, which is tied to the SSAMS common, i.e. the steel frame of the SSAMS, and by tweaking the generator voltage and phase outputs. We presumed that voltage drooping under load and subsequent compensation were causing transients in the common line, which were picked up by the EM discriminator. These results have yielded a lower noise threshold than demonstrated in Figure 17.

**Novel Primary Ion Beams and Sample Flooding Species**

In a later section we describe the design of our duoplasmatron arc supply. This arc supply is robust and stable, and has allowed us to explore several unconventional or little-used gas mixtures in the ion source. We have discovered several novel and useable ion beam species for depth profiling, imaging, and microanalysis, such as $NO_2^+$, $NO_2^-$ and $CFO^-$. There has been considerable literature published discussing the production of fluorinated ion beams [44-53], for instance, though none of these are routinely used today. A significant reason to use fluorine ion probes is that they typically enhance the yield of transition metals relative to oxygen probes. Many of these previous attempts have suffered from plasma instabilities and/or low ion beam intensities, which limit their routine utility. The parameter space of potential gas mixtures, duoplasmatron component compositions, and ion source operational parameters is extensive, and achieving a stable plasma is not trivial. A full discussion of these intricacies is beyond the scope of this paper, however Figure 18 demonstrates the utility of $NO_2$ for depth profiling analyses. This figure shows a comparison of $NO_2^+$, $NO_2^-$ and $O_2^-$ ion beams under different accelerating voltages to depth profile $^{55}Mn$ (55 keV, peak ~50 nm deep) and $^{60}Ni$ (60 keV) ion implants in Si metal. The presence of Fe at the surface indicates that some Fe was co-implanted and/or there is some Fe contribution from surface contamination. Closed symbols are depth profiles without any sample flooding and open symbols illustrate sample flooding with $CF_4$. Markers are thinned for visual clarity. Profiles were obtained by using 10 nA probes rastered over a 150 µm × 150 µm area with electronic gating for analysis of a central 50 µm × 50 µm region. Secondary ion energy was 4.5 keV, resulting in impact energies of, e.g., 3 keV for $NO_2^+$ (7.5 keV) and 14.5 keV for $NO_2^-$ (10 keV). $NO_2$ exhibits better ion yields, depth resolutions, and sputtering rates than $O_2$. Furthermore, the typical maximum ion current for $NO_2^+$ and $NO_2^-$ was 200 nA through the second primary beam aperture, making it a practical and robust ion probe for several types of analyses. While $O_2$ has been commonly used for sample flooding to enhance ion yields and/or smooth roughened crater bottoms [53-55], other non-oxygen flood gases have been observed to provide even higher secondary yields, e.g., $CCl_4$ [56]. We observe that $CF_4$ tends to enhance ion yields when they are already low (e.g., $NO_2^+$ and $O_2^-$), but only appears to affect the sputter rate when yields are higher (e.g., $NO_2^-$). This is not necessarily surprising given that there are limits to how efficiently a specific element will ionize, which also heavily depends upon the matrix, but it does open the door to renewed interest into research aimed at modifying surface sample chemistry via different combinations of primary beam and flood gas species. One of the main drawbacks of non-oxygen primary or flood species is that additional secondary molecule species are produced, which can confound isotope analyses[56]. The NAUTILUS, which universally dissociates all secondary molecules, does not suffer from this drawback, so any combination of gas species that enhances ion yields, or produces an enhanced and useful molecule for fragment mode analyses, will be useful and increase our sensitivity.

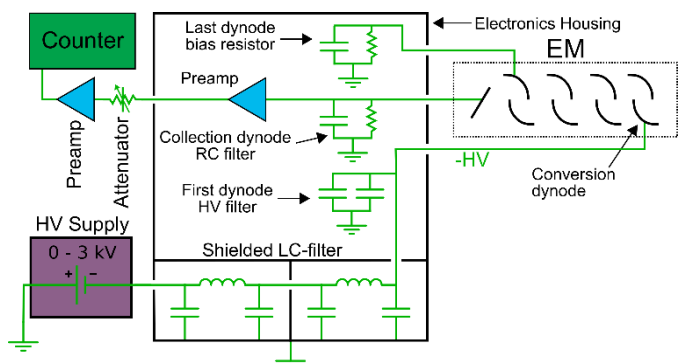

Figure 19: SIMS and SSAMS EM counting system schematic. We achieved pulse-counting deadtime of <10 ns.

### Instrument Control and Data Acquisition

The original SIMS and SSAMS instruments each came with their own commercial electronics and control computer. We unified control of the two systems onto one computer running in-house LabVIEW and Python software. Instrument control signals are provided by National Instruments PCI eXtensions for Instrumentation (PXI) hardware for digital waveforms and analog voltages, and a combination of TCP/IP, USB, GPIB, and RS-232/485 protocols to commercial off-the-shelf (COTS) components. Fiber optics and fiber-to-ethernet converters are used to communicate with instrumentation on the HV SSAMS. All of these modifications, including custom hardware and software, were required to achieve the NAUTILUS' demonstrated capabilities.

### Hardware

Instrument control and data acquisition hardware on the NAUTILUS is comprised primarily of COTS components. This allows for components to be upgraded and replaced in real time and for the NAUTILUS to remain at the forefront of current technologies. Low- and high-voltage commercial amplifiers and power supplies from several companies (Kepco, Inc. (Flushing, NY, USA); Trek, Inc. (Lockport, NY, USA); Spellman High Voltage Electronics Corporation (Hauppauge, NY, USA); TDK-Lambda (Tokyo, Japan); Spectrum Solutions, Inc. (Russellton, PA, USA); Sorensen/Ametek Programmable Power (San Diego, CA, USA); Heinzinger electronic GmbH (Rosenheim, Germany) are used to control ion optical components and magnet coils. Many of these power supplies amplify ±10 V, 16-bit analog control signals provided by National Instruments PXI(e) cards. Custom amplifier boards power the raster and dynamic transfer deflectors. Digital mixing boards made by Tangent, such as the Wave and Element series, are used for human interaction during instrument tuning. These mixing boards connect via USB to the computer. Each knob, button, and trackball is custom-defined to correspond to a lens, voltage-controlled element, or valve. Customizability is important for continued upgrading and improvement of the NAUTILUS.

### Pulse-Counting Electron Multipliers

We used ETP 14133H EMs for both high- and low-energy ion detection on the SIMS and SSAMS, respectively. Figure 19 shows a schematic of the pulse counting system, which provide <10ns deadtime on the SSAMS EM. One of the biggest sources of potential noise in the pulse counting configuration was RF pickup on the HV input lines. We prevented the HV LC filters from radiating to each other by placing each in a shielded compartment; electrical connections pass through apertures smaller than the RF waveguide cutoff size. Additionally, all seams on the electronics enclosure were sealed with aluminium tape to further prevent RF leaks. Discrete dynode EMs, like the 14133H, have their dynodes connected by a resistor chain from the biased conversion dynode down to ground. We connected the last dynode to ground through a parallel RC circuit to provide a small bias and extra charge for the amplified pulses. The collector dynode was connected to a Phillips Scientific 6950 amplifier, with an additional RC filter for pulse shaping and to prevent ringing. Outside of the electronics housing, we used a Phillips Scientific 5010 rotary attenuator and another 6950 amplifier to generate peak pulse heights of 0.5 V for counting with a Keysight 53230A counter/timer (50 Ω input impedance). A noticeable benefit of pulse-counting 304.5 keV ions on an EM is that we do not observe any detector mass fractionation, i.e. the efficiency at every mass appears to be nearly 100%.

### SIMS Magnet Control and Feedback

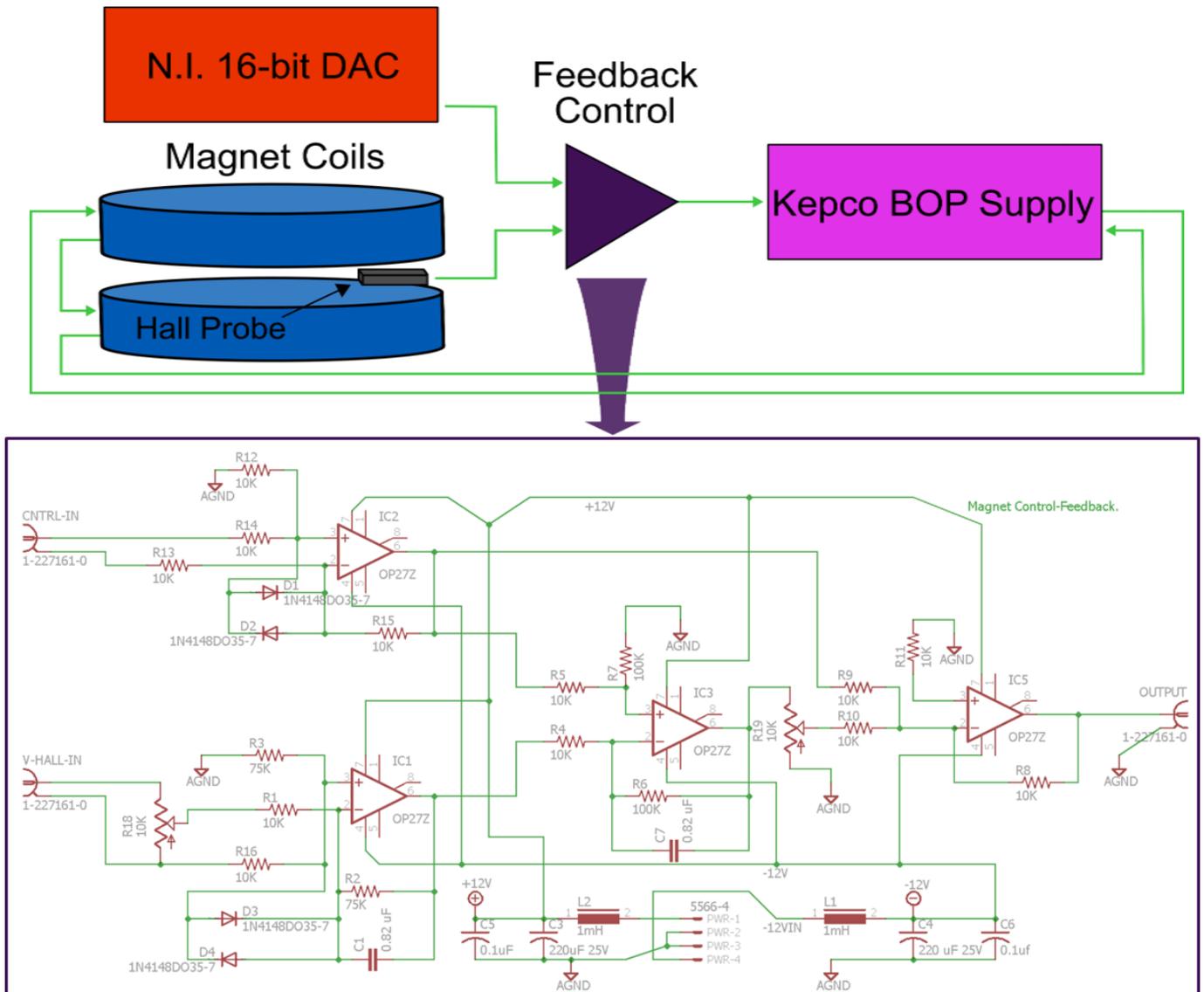

Figure 20: SIMS magnet control and feedback circuit. (Top panel) Block diagram where the feedback circuit sums the Hall probe analog output with the analog control signal from a DAC. (Bottom panel) Circuit diagram for the feedback system.

In order to switch the ims-4f magnetic field quickly and reproducibly we developed a simple control system that used Kepco Bi-polar operational power (BOP) supplies to provide the coil-current, a Group 3 Teslameter (Model DTM-151) to measure the field between the pole faces, and a 16-bit analog output from a National Instruments DAC as the control signal (16 bits is sufficient for low mass-resolution operation of the ims-4f magnet). The BOP supply was chosen to have a voltage and current output that most closely matched the DC resistance of the magnet coils. The Group 3 Teslameter has an analog output that was used in the feedback circuit to produce an error-signal to increase switching speed.

Without a feedback system to switch the ims-4f magnet it takes several seconds for the magnetic field to approach its equilibrium value. The long time constant for field-switching was due in part to the soft iron core of the magnet. With a feedback system in place the switching time was one second or less, depending on the relative size of the field change between peaks. A block diagram of the control system is shown in Figure 20 (top panel). The feedback circuit, shown in Figure 20 (bottom panel), takes the output signal from the Teslameter and produces an error signal with the control voltage in the differential amplifier (IC3 in the diagram). This error signal is summed in the final amplifier (IC5) to produce a control signal that drives the Kepco BOP supply. Values of the resistors are chosen to produce the appropriate matching signal levels and to choose the size of the error signal. Capacitive components in the feedback are chosen to reduce noise but also, most importantly, to reduce the bandwidth of the feedback so as to keep the system from oscillating. This is primarily accomplished with the final summing amplifier (IC5).

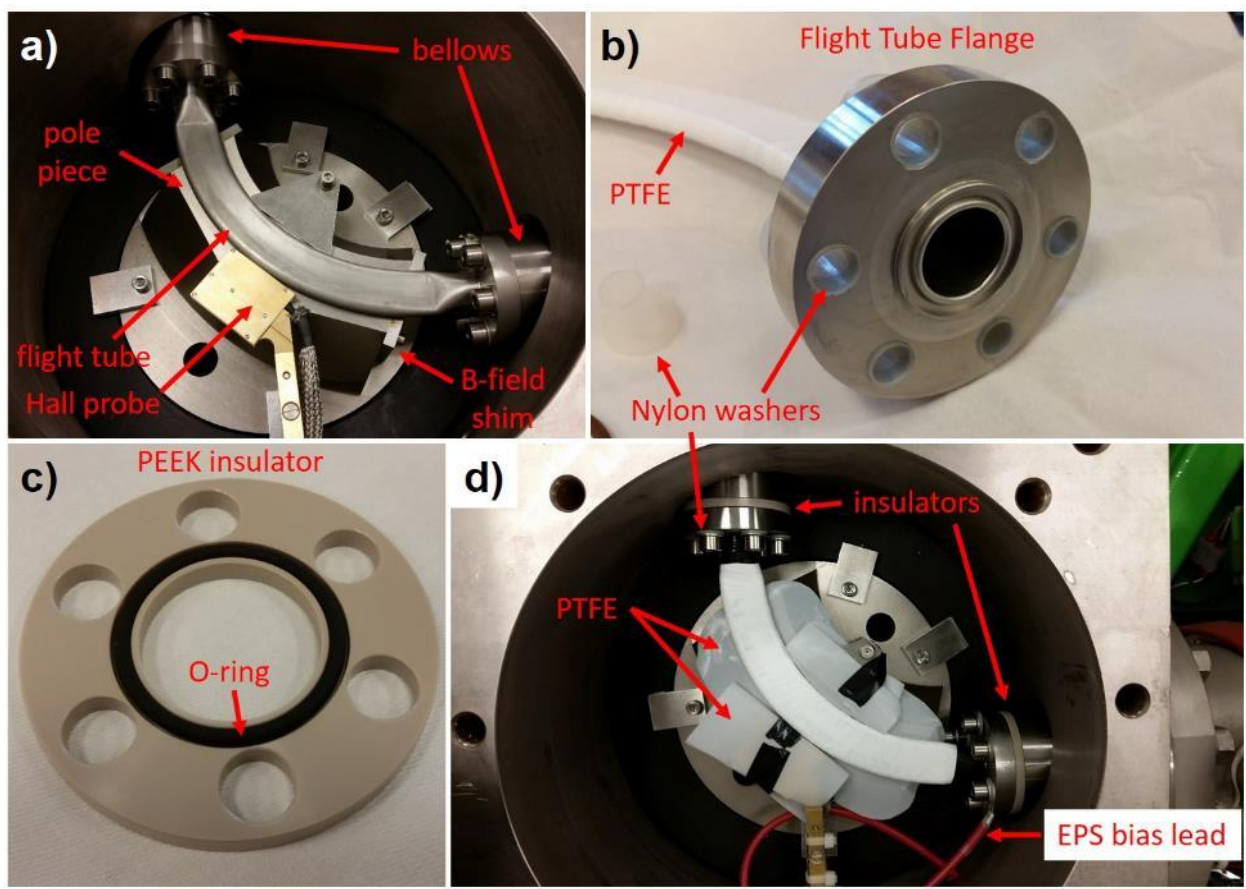

Figure 21: Modifications to enable EPS on the Cameca ims 4f SIMS magnetic sector. (a) Pre-modified magnetic sector with the top pole piece removed. (b) Flight tube wrapped in PTFE tape with insulating nylon shoulder washers. (c) Machined PEEK insulator with Viton O-ring. (d) Insulated flight tube installed with bias lead and extra PTFE sheeting on the pole piece and Hall probe holder. Schematics for the PEEK insulator are given in [6].

## Software

Instrument control software was written in house using a combination of National Instruments (NI) LabVIEW and Python implemented using Enthought's Python Integration Toolkit for LabVIEW (PITL). Instrument and hardware communication is performed using NI-DAQmx drivers and Virtual Instrument Software Architecture (VISA) through LabVIEW and Python (nidaqmx and pyvisa packages for the latter). In the main Tuning and Acquisition LabVIEW programs, multiple Python sessions are spawned using PITL, which allows data to be passed between them via TCP/IP. Number crunching tasks, such as real-time image analysis and waveform generation, are offloaded to the Python sessions, which provide a significant speed boost and lower overhead than LabVIEW alone. In addition, PITL is agnostic regarding the bitness of the Python sessions it spawns, so 32-bit or 64-bit Python interpreters may be called from 32-bit or 64-bit LabVIEW. This has allowed us to upgrade our LabVIEW installations to 64-bit versions, taking advantage of increased memory allocation, while still being able to run legacy software and drivers only available in 32-bit versions. This is accomplished by spawning a 32-bit Python process and using the ctypes module to call 32-bit dynamic-link libraries (dll). As of this writing, this workaround is used to communicate with our Tangent mixing boards (human interface devices, HID) used for tuning ion optics, and our sample stage stepper motors, which use the legacy NI Flexmotion architecture. We are in the process of gradually replacing LabVIEW with a more pure Python implementation of the software, though LabVIEW remains a useful tool for rapidly prototyping new programs.

Data are saved in Hierarchical Data Format (HDF5)[57] files. In addition to being open source, HDF5 files are self-describing and can contain arbitrary heterogeneous data, such as images, waveforms, and simple strings and numbers. There exist HDF5 wrappers for nearly all major programming languages, including C, Python, Java, and others, making the data portable and easy to share between laboratories. In-house Python GUIs and scripts are used for data extraction, and image and depth-profile analysis.

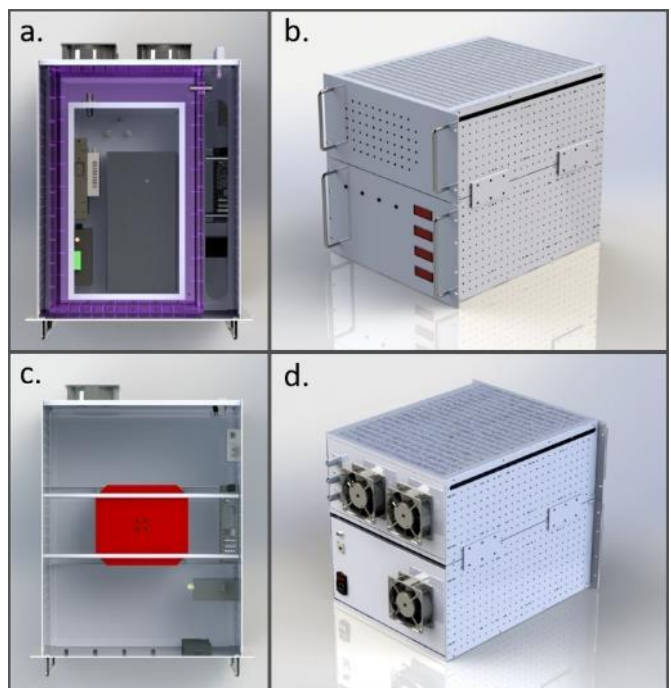

Figure 22: 3D CAD rendering of duoplasmatron arc supply. The arc supply is split between low-voltage (lower) and high-voltage (upper) chassis. Top view of the HV chassis is shown in panel a, with the insulating Lexan box shown in purple. Top view of the LV chassis shown in panel c, with the isolation transformer in red. Circuit diagram shown in Figure 23.

### Cameca ims 4f Modifications

The entirety of the electronics and non-vacuum system hardware on the ims 4f have been replaced, except for the ion pump control unit. This includes all power supplies for lenses, deflectors, ion sources, and magnets; turbo pumps; hall probes; and detector counting systems. As many of the power supplies and amplifiers as possible are COTS components controlled by analog or digital means. Most mono-polar HV lens supplies have their outputs set in parallel to grounded bleeder resistors, nominally for 10% of the maximum power output, for stability and more rapid current sinking. As described earlier in the text, the SIMS EM and FC detectors were moved into a cube outside of the original SIMS vacuum system and are positioned off-axis from the path that leads into the accelerator. Deflectors are used to select each detector, and the projection lenses are used to compensate for the difference in detector distance from the original mass spectrometer layout.

The design of the duoplasmatron arc supply is given later. In addition to the new arc supply, we fabricated a semi-magnetic split anode for the duoplasmatron following the description in Williams et al.[58]. This design significantly boosts the negative ion beam yields from the duoplasmatron since it prevents electrons from being extracted from the plasma, which often causes the bias voltage to droop. With this modification, the intermediate or z-electrode can also be positioned nearly at the center of its motion for maximum beam intensity.

### Electrostatic Peak Switching (EPS)

Our Cameca ims 4f was originally equipped with an insulated magnet flight tube for EPS, however this feature was removed prior to its arrival at NRL. We re-insulated the flight tube by fabricating a set of polyether ether ketone (PEEK) annuli with an O-ring groove on one side to place between the bellows and flight tube flanges. Nylon shoulder washers were used to insulate the bolts. Figure 21 shows a view of the 4f magnetic sector, with the top pole piece removed, before and after installation of the insulating components. PTFE sheets and tape were used to insulate the flight tube from the pole pieces and Hall probe. A schematic of the PEEK insulator with dimensions can be found in Groopman et al.[6].

## Duoplasmatron Arc Supply

We designed and built a robust duoplasmatron arc supply that can be operated digitally via any computer with an Ethernet connection. A 3D CAD model of the supply and circuit diagrams for its construction are shown in Figure 22 and Figure 23, respectively. Since there are many legacy SIMS instruments still in use globally, it is our hope that these designs may be of use to research groups looking to extend the life of and improve the capability of their instruments. The authors may be contacted directly for parts lists, CAD models, and schematics.

In brief, the arc supply consists of a regulated, current-control HV power supply that is floated at the duoplasmatron extraction potential. We chose a Spellman SLM -3kV 100mA (300 W) power supply to provide the arc current to the duoplasmatron cathode. A bipolar 15 kV Trek supply is used as our bias. In the upper, HV chassis (Figure 23), a Lexan isolation box houses an isopotential aluminum Bud Box, which houses the HV electronics and is floated by the bias supply. AC power is provided by an isolation transformer in the lower, LV chassis. This provides 120 VAC to the SLM supply and a 24/5 VDC regulated power supply, which powers communication devices. For communication, a fiber optic line connects two fiber/Ethernet converters, one in each chassis, which digitally controls the SLM via Ethernet. The Ethernet connection provides queried status updates and outputs to the computer in addition to power supply control. We also used an analog/fiber converter to transport the analog monitor voltages from the SLM to ground, where they are scaled and displayed in real-time on digital panel meters on the chassis. The SLM output in connected to the duoplasmatron cathode in parallel to a 66.4 kΩ bleeder resistor, which helps stabilize the reactive plasma. All of the components in the arc supply itself are COTS, and therefore easily replaceable. For instance, Spellman also offers and 600 W version of the SLM (200 mA) with the same form factor, which could be easily substituted.

In addition to the arc supply, we modified our duoplasmatron by replacing the anode with a semi-magnetic version developed and shared by Peter Williams and Richard Hervig[58]. This anode reduces the extracted electron current during negative ion beam generation, which can cause voltage drooping in the bias supply and in our case often resulted in arcing when the duoplasmatron's intermediate electrode was aligned with the extraction axis. This anode design enhanced our plasma stability when generating non-O negative ion beams.

## NEC SSAMS Modifications

We have made several updates to the NEC SSAMS system following its delivery. Crucial upgrades to the SSAMS included the addition of two MCPs for beam imaging, one on a linear motion feedthrough immediately following the gas stripping cell and one at the end station on the ESA focal plane with the EM; the two end station detectors are switched between electrostatically. Being able to tune an image of the beam at several locations was significantly more expedient than attempting to maximize detector counts, since the secondary lens settings of the SIMS are non-standard and the optimal settings for the SSAMS were unknown. The channel plates additionally aided in identifying the cause of a mass fractionation issue present in Fahey et al.[7], which we have since rectified. As discussed previously, we also made several small modifications to reduce the electronic noise on our detectors, which including installing better isolation between the Hall probe and floating flight tube in the magnetic sector, and by adding TVS diodes to the AC generator legs to reduce transients.

## Elimination of Mass Fractionation from Fahey et al. (2016)

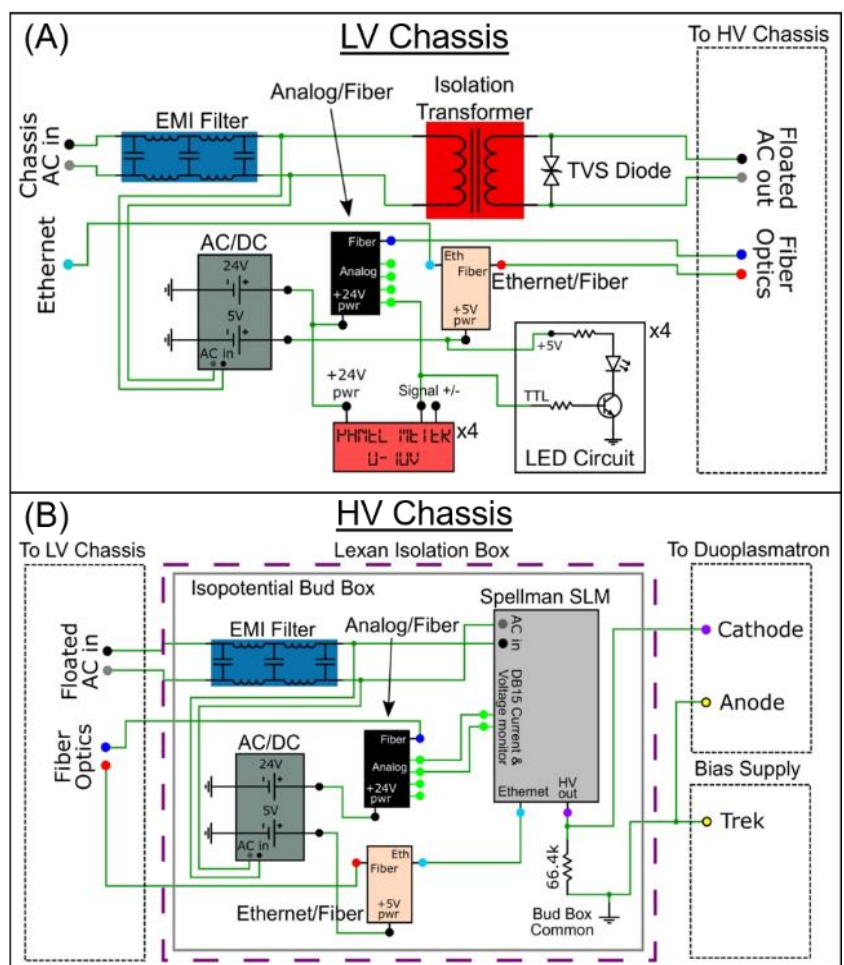

Figure 23: Circuit diagram for duoplasmatron arc supply. The supply is split between two chassis, one containing low-voltage electronics (panel A), and the other containing the floated high-voltage electronics (panel B).

In 2016 we found mass-dependent fractionation in U isotope analyses of NIST Certified Reference Material (CRM) particles[7]. In that article we corrected the minor isotope ratios using the $^{235}U/^{238}U$ ratio and inferred that we had a systematic error in our peak centering routine for the minor isotopes. Based upon the schematics provided by NEC and SIMION simulations following publication, we discovered that a gap lens attached to the floating SSAMS magnet flight tube was installed backwards, which we rectified, removing the mass bias. The circular gap lens is located nearly in line with the flange on one end of the ~6.5" long insulated nipple in which it is housed. Figure 24 shows a schematic cross section of the gap lens at the entrance of the SSAMS magnet flight tube (mirrored on the exit), and the corresponding SIMION models with field lines for the original and current orientations of the lens. When the gap lens was proximate to the rectangular entrance to the magnet flight tube, a quadrupole-like focusing effect occurred when the flight tube was biased, stretching the beam vertically for one polarity and horizontally for the other (Figure 25). This resulted in an EPS dependent (i.e. mass) effect on the isotope ratios in Fahey et al.[7], since the defocusing of the beam out of the detector collection area scaled with the magnitude of the flight tube bias. From the SIMION model, it is clear that the field lines are not symmetric for the top and side views in the original orientation (Figure 24b,d). Since the gap lens-flight tube system is a non-ideal Einzel lens[6], we expect some steering and focusing effects that are asymmetric with the bias voltage, but these are small, as seen in the bottom panels of Figure 25. We now find no measureable mass fractionation from the EPS setup in our measurements and our analytical uncertainties are predominantly driven by counting statistics and spot-to-spot inhomogeneities, in line with other SIMS instruments.

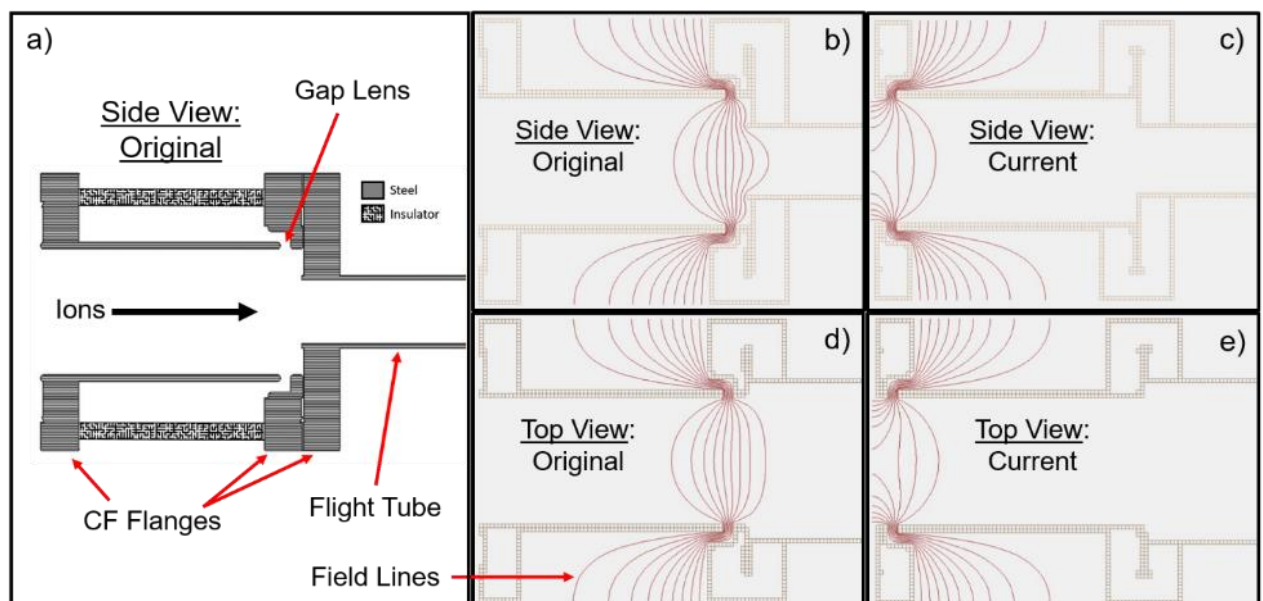

Figure 24: Schematic and SIMION models of the gap lens at the entrance and exit of the SSAMS magnetic flight tube. The original installation and NEC schematics of the gap lens had the gap proximate to the rectangular entrance of the flight tube, causing a quadrupole-like focusing effect and mass-dependent fractionation seen in [7]. SIMION modelling coupled with measurements helped identify the source of the mass fractionation, which was fixed by reversing the orientation of the gap lens.

## Conclusions

We have built a novel mass spectrometer at the U.S. Naval Research Lab, NAUTILUS, which combines a full SIMS instrument with a molecule-filtering SSAMS "detector". The NAUTILUS addresses problems that lie at the nexus between the traditional boundaries of SIMS and AMS. SIMS excels at *in situ*, spatially resolved surface analyses, though nuclear and molecular isobars can interfere with specific measurements; AMS excels at ultra-trace isotope measurements with high dynamic range, however preparatory chemistry or sizeable sample requirements eliminate micro-scale petrologic context. The NATUILUS provides complementary capabilities to both techniques. Of particular note is the NAUTILUS' ability to collect molecule-free raster ion images for rapid analysis of trace elements in complex, heterogeneous matrices. In complex matrices, such as spent nuclear fuel, petrologic context is incredibly important, but spot-to-spot matrix and isotopic heterogeneities challenge molecule or isobar corrections. Direct isotope imaging is similarly of great utility for particle searching based upon a specific isotopic signature, again without the need to make corrections based upon other isotopes. By eliminating the molecular background, which is omnipresent in SIMS, the NAUTILUS further takes advantage of novel primary ion species and/or sample flooding gases, which otherwise complicate the speciation of the secondary molecular ions. Any boost to an atomic or molecular secondary ion signal is useful to the NAUTILUS' sensitivity.

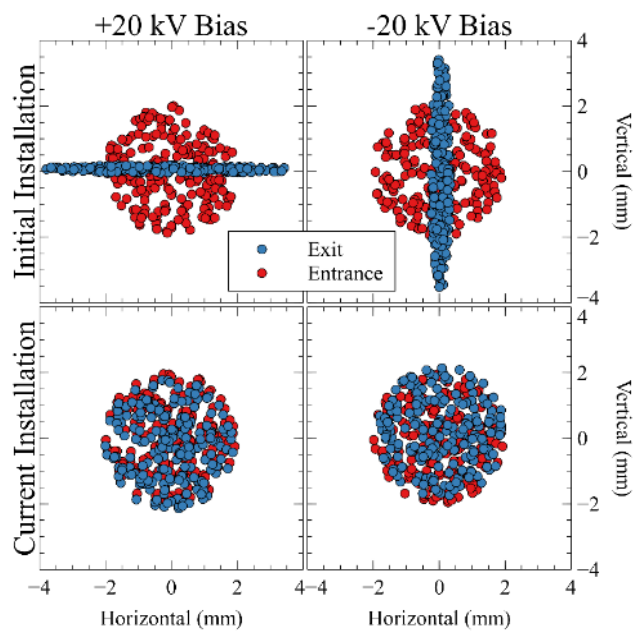

Figure 25: Cross sectional comparison of a cylindrical beam injected through the SSAMS flight tube and gap lenses performed in SIMION. Reversing the gap lens direction removed the coupling of field lines between the circular gap lens and the rectangular flight tube entrance.

The NAUTILUS also achieves high stability and day-to-day reproducibility by substituting low MRP and molecular dissociation for high MRP used by conventional SIMS. The ultra-low background and quiet detector electronics maximizes the sensitivity of the NAUTILUS to trace element analyses from micro-scale volumes of material, which are far too small to be probed by other AMS techniques. These developments have found immediate application in several fields since they are targeted at otherwise difficult, if not intractable problems, for both of NAUTILUS' parent techniques.

### Acknowledgements

Funding for the NAUTILUS was provided by the Office of Naval Research through NRL via capital purchases and the Base Program. We thank Victor Cestone, Kamron Fazel, and David Knies for their contributions during development. We thank Alex Meshik and Olga Pravdivtseva (WashU), Maurice Pagel, and Francois Gauthier-Lafaye for Oklo sample; the National Museum of Natural History (NMNH) for the monazite samples; Allen Kennedy for the Grenville skarn titanite sample; Laure Sangely (IAEA) for uranium particles; Amy Jurewicz (ASU) and Donald Burnett (CalTech) for Mn, Ni ion implants. We thank Peter Williams and Rick Hervig (ASU) for the split anode design and discussion. We thank Bernie LaFrance for machining split anode, PEEK insulators, and duo parts.